\documentclass[reprint,amsmath,amssymb,aps, pra, superscriptaddress,nofootinbib
]{revtex4-2}

\usepackage{graphicx}
\usepackage{dcolumn}
\usepackage{bm}
\usepackage{xcolor}
\usepackage{hyperref}

\hypersetup{
	colorlinks = true,
	linkcolor = blue,
	citecolor = blue,
	urlcolor = blue,
}

\begin{document}
	
	\preprint{APS/123-QED}
	
	\title{Cavity Optomechanical Quantum Memory for Twisted Photons Using a Ring BEC}
	
	\author{Nilamoni Daloi}
	\email{nilamoni123@gmail.com}
	\affiliation{School of Physics and Astronomy, Rochester Institute of Technology, Rochester, New York 14623, USA}
	
	\author{Rahul Gupta}
	\affiliation{Department of Physics, Indian Institute of Technology Bombay, Mumbai 400076, India}
	
	\author{Aritra Ghosh}
	\affiliation{School of Basic Sciences, Indian Institute of Technology Bhubaneswar, Odisha, 752050, India}
	
	\author{Pardeep Kumar}
	\affiliation{Max Planck Institute for the Science of Light, Staudtstraße 2, 91058 Erlangen, Germany}
	
	\author{Himadri S. Dhar}
	\affiliation{Department of Physics, Indian Institute of Technology Bombay, Mumbai 400076, India}
	\affiliation{Centre of Excellence in Quantum Information, Computation, Science and Technology, Indian Institute of Technology Bombay, Mumbai 400076, India}
	
	\author{M. Bhattacharya}
	\affiliation{School of Physics and Astronomy, Rochester Institute of Technology, Rochester, New York 14623, USA}

	\begin{abstract}
		We theoretically propose a photonic orbital angular momentum (OAM) quantum memory platform based on an atomic Bose-Einstein condensate confined in a ring trap and placed inside a Fabry–Pérot cavity driven by Laguerre-Gaussian beams. In contrast to electromagnetically induced transparency–based protocols, our memory does not require change of internal atomic levels. The optical states are instead stored in the large Hilbert space of topologically protected and long-lived motional states (persistent currents) of the condensate, yielding a storage time three orders of magnitude better than presently available. Further, the use of a cavity provides orders of magnitude more resonances, and hence bandwidth, for reading and writing than internal atomic transitions. Finally, the analogy to cavity optomechanics suggests a natural path to wavelength conversion, OAM transduction, and nondestructive readout of the memory.  
	\end{abstract}
	\maketitle
	%
	%
	%
	%
	%
	%
	%
	%
	
	\section{\label{sec:Introduction}Introduction}
	Photonic quantum memories~\cite{Heshami12112016, lvovsky_optical_2009, veissier:tel-00977307} are essential components of quantum information networks, designed to store and retrieve quantum states of light on demand. In recent years, the orbital angular momentum (OAM) of photons has attracted significant attention as a high-dimensional basis for encoding quantum information, owing to its unbounded Hilbert space~\cite{he_towards_2022}. Existing quantum memory protocols for OAM-encoded photons have primarily relied on cold atomic ensembles~\cite{Shi_2018} via electromagnetically induced transparency (EIT)~\cite{ding_high-dimensional_2016, PhysRevLett.114.050502, nicolas_quantum_2014, PhysRevA.79.023825, Veissier:13, Wang_2021}, or on rare-earth-ion-doped crystals~\cite{PhysRevLett.115.070502}. The longest reported storage time for a single photon carrying OAM is $400$ $\mu$s, achieved using a cold rubidium atomic ensemble~\cite{PhysRevLett.129.193601}.     
	
	In this context, it is crucial to note that a Bose-Einstein condensate (BEC)~\cite{leggett_bose-einstein_2001} confined in a ring-shaped optical trap~\cite{PhysRevA.74.023617, PhysRevA.83.043408} supports quantized rotation in the form of persistent currents~\cite{PhysRevLett.99.260401, PhysRevLett.110.025302, PhysRevLett.124.025301}. These circulating matter waves are topologically protected and metastable, and hence exhibit long lifetimes lasting up to a minute~\cite{PhysRevLett.106.130401,PhysRevLett.110.025301}. Thus, mapping photonic quantum states onto such atomic persistent current modes presents a promising route toward realizing long-lived OAM quantum memories. Earlier theoretical work on such a possibility, limited to OAM superposition states, relied on internal atomic levels for storage and retrieval in the degenerate gas~\cite{kapale_dowling_vortex_phase_qubitPRL2005}. The reliance on internal atomic levels puts severe constraints on the memory, since only an order of unity (ten) hyperfine (Zeeman) ground states are typically available for trapping, condensation, and storage, and only a handful of laser-accessible atomic transitions can be used for reading and writing.

	In this paper, we propose a quantum memory platform that removes these obstacles and also offers a storage time three orders of magnitude longer than currently available: a cavity ring BEC system for the storage and retrieval of photonic states encoded in the OAM basis of light. Rather than involving internal atomic states (in our scheme all atoms stay in the same internal state throughout the protocol), we store information purely in the \textit{motional} states of the atoms, i.e. in the persistent currents. A large number of such states are available, with winding numbers $\sim 10^{4}$ having been realized experimentally~\cite{PhysRevLett.124.025301,von_Klitzing_hypersonic_BEC}. Also, rather than relying on a small number of atomic transitions occurring at wavelengths specified by nature, for writing and storage, we propose the use of the optical resonances of a cavity. These are much larger in number than atomic transitions and can be easily tuned by changing the cavity length, and thus allow significantly more bandwidth for reading, writing, wavelength conversion~\cite{PhysRevA.82.053806} and OAM transduction~\cite{Kaviani:20}. For example, in a $100$ nm range in the visible, there are $\sim 10^{5}$ equally spaced resonances for a centimeter-long cavity. Specifically, our platform supports storage of single-photon Fock states carrying OAM of either sign~\cite{PhysRevLett.88.257901}, coherent superpositions of counter-rotating OAM states~\cite{PhysRevLett.129.193601}, and entangled two-photon states with equal and opposite OAM~\cite{Shi_2018}. For realistic experimental conditions, we show that storage times reaching hundreds of milliseconds are achievable, representing three orders of magnitude improvement over existing capabilities~\cite{PhysRevLett.129.193601}.
	%
	%
	%
	%
	%
	%
	%
	%
	\begin{figure}[ht]
		\centering
		\includegraphics[width=\linewidth]{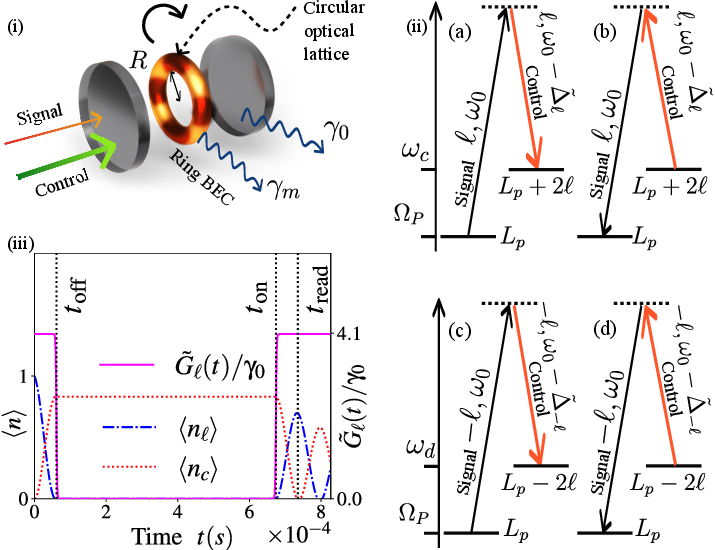}
		\caption{\label{fig:figure1}(i)~BEC rotating in a ring trap of radius $R$, inside a Fabry-P\'erot cavity, driven by control laser signal fields each in a coherent superposition of LG modes carrying OAM $\pm\ell\hbar$. (ii)(a,b) and (c,d) show the energy and OAM selection rules for storage and retrieval of a single signal photon with OAM $+\ell\hbar$ and $ -\ell\hbar$, respectively. (iii)~Read and write protocol for $\ell\hbar$ OAM mode. Parameters used for (iii): control power$\sim8.6\times10^{-7}$ mW, $R = 10$ $\mu$m, $L_p = 20$, $\ell = 130$, $N = 2\times10^4$, $\gamma_0/2\pi = 1$ kHz, $\gamma_m/2\pi \sim 1.7\times 10^{-5}\gamma_0$, $\tilde{G}/2\pi = 4\gamma_0$, $t\textsubscript{off} = \pi/2\tilde{G} \sim 61$ $\mu$s, $t\textsubscript{on} \sim 674$ $\mu$s, $t\textsubscript{read} \sim 735$ $\mu$s.}
	\end{figure}
	%
	
	
	\section{\label{sec:Theoretical_model}Theoretical model}
	
	We consider a BEC of $N$ identical two-level $^{23}\mathrm{Na}$ atoms of mass $m$, confined in a ring trap~\cite{PhysRevA.74.023617} of radius $R$, placed inside a Fabry-P\'erot cavity of length $L$, resonance frequency $\omega_0$, and photon decay rate $\gamma_0$, as illustrated in Fig.~\ref{fig:figure1}(i). The cavity is driven through one of the mirrors by two optical fields; a strong classical control laser with frequency $\omega_L$ and pumping rate $\varepsilon_c$ and a weak quantum signal laser of frequency $\omega_p$ and pumping rate $\varepsilon_p$. Both fields are prepared in coherent superpositions of Laguerre-Gaussian (LG)~\cite{Yao:11} modes carrying OAM $\pm \ell\hbar$, generating a circular optical lattice about the cavity axis that overlaps with the ring-shaped BEC as shown in Fig.~\ref{fig:figure1}(i). The BEC undergoes quantized rotational motion around the cavity axis, characterized by a winding number $L_p \in \mathbb{Z}^{+}$~\cite{PhysRevLett.110.025302}. The associated rotational energy is given by $\hbar\Omega_p$, where $\Omega_p = \hbar L_p^2/2I$~\cite{kumar_cavity_2021} with $I=mR^{2}$ being the moment of inertia of an atom of mass $m$ about the cavity axis.

	The optical lattice formed by the control field causes Bragg scattering of atoms in the BEC from their initial rotational mode with winding number $L_{p}$ to side modes with winding numbers $L_p \pm  2n\ell$ ($n =  1, 2,\cdots$). The optical fields are blue detuned far from the atomic resonance, making the light-atom interaction very weak. Thus, only first-order Bragg scattering ($n = 1$) is appreciable, and the majority of atoms remain in the mode $L_p$. Under these conditions, the system Hamiltonian in the rotating frame of the control field is given by~\cite{kumar_cavity_2021},

	\begin{align}
		\label{eq:ring_BEC_Hamiltonian_1}
		H/\hbar &= - \tilde{\Delta}a^\dagger a + \omega_c c^\dagger c + \omega_d d^\dagger d + G(X_c + X_d)a^\dagger a \notag\\
		& + i(\varepsilon_ca^\dagger - \varepsilon^*_c a) + i(\varepsilon_p a^\dagger e^{-i\delta t} - \varepsilon^*_p a e^{i\delta t})\notag\\
		& + 4\tilde{g}N(c^\dagger c + d^\dagger d) + 2\tilde{g}N(c d + c^\dagger d^\dagger).
	\end{align}
	Here, $a$ denotes the annihilation operator for the intra-cavity field, while $c(d)$ and $X_{c(d)}$ represent the annihilation and position quadrature operators, respectively, for the BEC side mode with winding number $L_p + (-) 2\ell$. The energy of the atoms in the mode $c(d)$ is given by $\hbar\omega_{c(d)}$ where $\omega_{c(d)} = \hbar[L_p + (-) 2\ell]^2/2I$~\cite{kumar_cavity_2021}. The parameter $G$~\cite{kumar_cavity_2021,aspelmeyer_cavity_2014} denotes the effective optomechanical coupling rate. For the parameters considered in this work, the effective detuning of the control laser from cavity resonance is given by $\tilde{\Delta} \approx \omega_L - \omega_0$. The two-photon detuning is defined as $\delta = \omega_p - \omega_L$. The final two terms in the Hamiltonian account for interatomic interactions, with interaction strength characterized by $\tilde{g}$~\cite{kumar_cavity_2021}. For typical parameters, the effect of weak interatomic interactions is small and Eq.~\eqref{eq:ring_BEC_Hamiltonian_1} reduces to the canonical optomechanical Hamiltonian~\cite{kumar_cavity_2021, aspelmeyer_cavity_2014}. The effects of atomic interactions have been briefly discussed in Appendix~\ref{App-A}.
	%
	%
	%
	%
	%
	%
	%
	%
	
	\section{Photons with OAM $+\ell\hbar$}
	
	The procedure to store a single-photon Fock state with OAM $+\ell\hbar$ in the cavity-BEC setup of Fig.~\ref{fig:figure1}(i) follows the energy level diagrams in Figs.~\ref{fig:figure1}(ii)(a-b) and implements the standard optomechanical memory protocol~\cite{fiore_storing_2011}, extended to conserve OAM. The control laser, with OAM $-\ell\hbar$ and frequency $\omega_L$, is red detuned from the cavity resonance $\omega_0$ by the frequency gap between the two BEC modes with winding numbers $L_p$ and $L_p + 2\ell$, respectively. This detuning is given by $\tilde{\Delta}_\ell = -(\omega_c - \Omega_p) \approx -\omega_c$, as $\Omega_p \ll \omega_c$ for typical parameters used in this work. The signal laser, carrying OAM $+\ell\hbar$ and at frequency $\omega_p$ is tuned to cavity resonance $\omega_0$, with the two-photon detuning $\delta = \omega_p - \omega_L = \omega_c+4\tilde{g}N$. For the chosen parameters, $\omega_{c}\gg 4\tilde{g}N$, making $\delta \approx \omega_{c}$. At the end of storage, the signal photon quantum state is transferred to a long-lived phonon of energy $\hbar\omega_c$ that serves as the memory. To retrieve the stored state, the control laser is turned back on with the same red detuning, thereby absorbing the phonon and regenerating the signal photon. 
	Similarly, storage and retrieval of a signal photon with OAM $-\ell\hbar$ require the phonon modes $L_p$ and $L_p - 2\ell$, and a control photon with OAM $+\ell\hbar$ [Figs.~\ref{fig:figure1}(i)(c,d)]. In this case, the control field detuning is given by $\tilde{\Delta}_{-\ell} = -(\omega_d - \Omega_p)\approx -\omega_d$ and the two-photon detuning by $\delta = \omega_d$. 
	
	Next, we outline the modifications to the Hamiltonian in Eq.~\eqref{eq:ring_BEC_Hamiltonian_1} required to implement the above protocol. In the weak signal limit ($\varepsilon_c\gg\varepsilon_p$) and the resolved sideband regime ($\omega_{c(d)}\gg \gamma_0$), the intra-cavity field mode can be decomposed as $a = \alpha + \tilde{a} e^{-i\delta t}$~\cite{doi:10.1126/science.1195596, PhysRevA.107.013525} where, $\alpha = \langle a\rangle$ denotes the classical amplitude of the strong control field and $\tilde{a}$ the quantum signal field mode. Similarly, the atomic sidemodes are expressed as $c =  \beta_c + \tilde{c} e^{-i\delta t}$ and $d =  \beta_d + \tilde{d} e^{-i\delta t}$, where $\beta_{c(d)} = \langle c(d)\rangle$. To suppress crosstalk between $c$ and $d$, we further impose $|\omega_c - \omega_d|\gg\gamma_0$, allowing adiabatic elimination of the off-resonant BEC mode. Under these conditions, the Hamiltonian in Eq.~\eqref{eq:ring_BEC_Hamiltonian_1} for storing the $+\ell\hbar$ OAM state linearizes to
	\begin{align}
		H_\ell/\hbar &= -\Delta^\prime\tilde{a}^{\dagger}\tilde{a} + \omega_c \tilde{c}^{\dagger}\tilde{c} + (\tilde{G}\tilde{a}^{\dagger}\tilde{c} + \tilde{G}^*\tilde{a} \tilde{c}^{\dagger}),\label{eq:single_mode_linearized_Hamiltonian}
	\end{align}
	where $\tilde{G}(t) = \alpha(t) G/\sqrt{2}$ is the equivalent boosted optomechanical coupling constant and $\Delta^\prime = \tilde{\Delta}_\ell - G\sqrt{2}\Re[\beta_c + \beta_d]$ is an effective detuning. For typical parameters, the second term of $\Delta^\prime$ is negligible ($\Delta^\prime\approx\tilde{\Delta}_\ell$). For brevity, we denote $\tilde{a}$, $\tilde{c}$, and $\tilde{d}$ as $a$, $c$, and $d$, respectively. Without loss of generality, we fix the phase of the control field such that $\bar{\alpha}\in \mathbb{R}$, and consequently, $\tilde{G}\in\mathbb{R}$. Under transformation, $U = \exp\left[-i\delta t a^\dagger a\right]\otimes \exp\left[-i\delta t c^\dagger c\right]$ and $\tilde{\Delta}_\ell = -\delta$, Eq.~\eqref{eq:single_mode_linearized_Hamiltonian} reduces to $H_\ell \sim \hbar\tilde{G}(a^{\dagger} c + a c^{\dagger})$. For the conditions discussed above, only the $+\ell\hbar(-\ell\hbar)$ OAM component is transferred to the mode $c(d)$. Thus, we may write, $H_\ell \sim  \hbar\tilde{G}(a^{\dagger}_\ell c + a_\ell c^{\dagger})$.
	
	For numerical simulation, we solve the following Lindblad equation~\cite{Quantum_Noise}

	\begin{align}
		\label{eq:master_equation}
		\dot\rho=\frac{1}{i\hbar}[H_\ell,\rho] -\frac{\gamma_0}{2}\mathcal{D}\left[a_\ell\right]\rho -\frac{\gamma_m}{2}\mathcal{D}\left[c\right]\rho,
	\end{align}
	where $\rho$ is the density matrix for the cavity-BEC system, $\gamma_m$ is the damping rate of the atomic mode, and $\mathcal{D}[\mathcal{O}]\rho = \{\mathcal{O}^\dagger \mathcal{O},\rho\} - 2\mathcal{O}\rho \mathcal{O}^\dagger$. After solving Eq.~\eqref{eq:master_equation}, the density matrix of the optical signal is obtained by tracing out the BEC mode. The time dynamics of the simulation for writing and reading the state $|1\rangle_\ell$ is shown in Fig.~\ref{fig:figure1}(iii). Since $H_\ell$ resembles a beam-splitter-type interaction, it coherently swaps the quantum states of modes $a_\ell$ and $c$ with a period $\pi/2\tilde{G}$. Thus, a state initialized in $a_\ell$ at $t = 0$ is transferred to $c$ after that duration. Turning off the control field at this time ($t\textsubscript{off}$) effectively stores the quantum state in $c$ [Fig.~\ref{fig:figure1}(iii)]. Optimal readout is achieved by reactivating the control field at $t\textsubscript{on}$ and measuring the output at $t\textsubscript{read} = t\textsubscript{on} + \pi/2\tilde{G}$.
	
	\begin{figure}[t!]
		\centering
		\includegraphics[width=\linewidth]{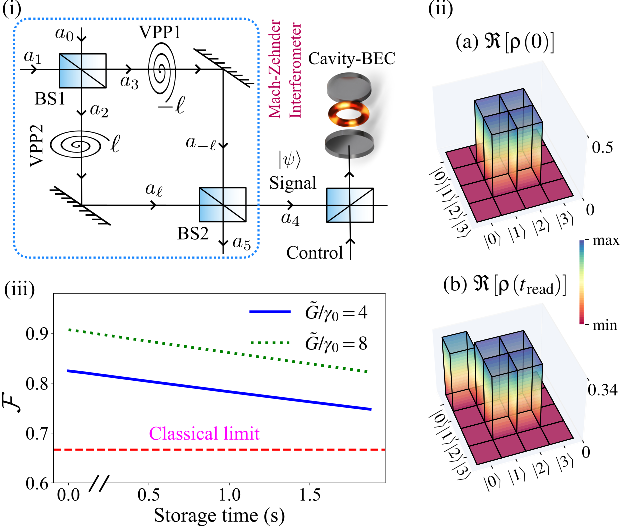}
		\caption{\label{fig:figure2}(i)~Schematic for generating an OAM superposition state. The beamsplitters and vortex phase plates are denoted by BS and VPP, respectively. The optical modes are denoted by $a_{i}, i\in (0-5)$. (ii)(a)~Initial density matrix for $|\psi\rangle$ of Eq.~\eqref{eq:OAM_superposition_state}, (ii)(b)~Retrieved density matrix. (iii)~Retrieval fidelity ($\mathcal{F}$) versus storage time ($\sim 0.6$~$\mu$s to $1.9$ s) for different values of $\tilde{G}/\gamma_{0}$. The red dashed line at $2/3$ in (iii) shows the classical limit of fidelity. Parameters for (ii) are the same as in Fig.~\ref{fig:figure1}.}
	\end{figure}
	%
	%
	%
	%
	%
	%
	%
	%
	\begin{figure*}[t]
		\centering
		\includegraphics[width=\linewidth]{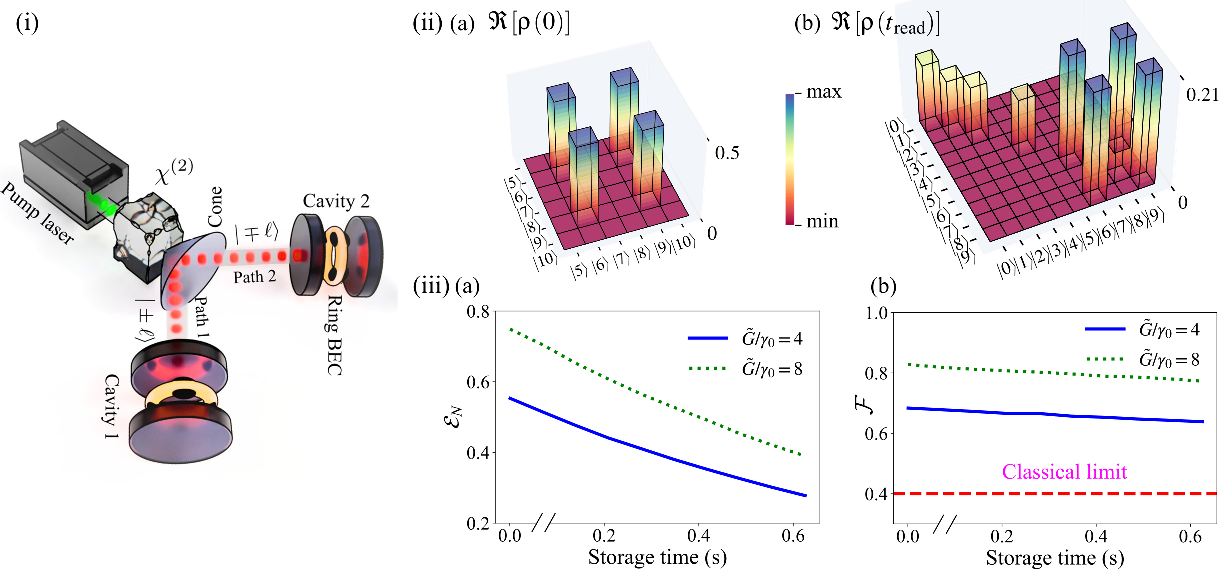}
		\caption{\label{fig:figure3}(i)~Schematic diagram for storage and retrieval of OAM entangled state. (ii)(a,b)~Intial and retrieved density matrices. (iii)(a)~Logarithmic negativity $\mathcal{E}_N$ versus storage time ($\sim 0.6$~$\mu$s to $0.6$ s). (iii)(b)~Corresponding retrieval fidelity ($\mathcal{F}$) versus storage time for various $\tilde{G}/\gamma_{0}$. The red dashed line denote the classical fidelity limit. Parameters for (ii) are the same as in Fig.~\ref{fig:figure1}.}
	\end{figure*}
	
	\textit{Superposition of OAM $\pm\ell\hbar$ states}.—We now consider the storage of a coherent superposition of the OAM state of light given by $|\psi\rangle = \left(|\ell\rangle + |-\ell\rangle\right)/\sqrt{2}$. This state can be created using a Mach-Zehnder interferometer as shown in Fig.~\ref{fig:figure2}(i); a brief analysis of the scheme~\cite{Galvez_2011} is provided in the Appendix~\ref{App-B}. In our scheme, two control fields with detunings $\tilde{\Delta}_{\ell(-\ell)} =-\omega_{c(d)}$ are used to mediate the storage of $|\ell(-\ell)\rangle$. The linearized Hamiltonian governing the dynamics is given by
	\begin{align}
		\label{eq:OAM_superposition_Hamiltonian}
		H\textsubscript{sup}/\hbar  &= \tilde{G}_\ell(a_{\ell}^\dagger c + a_{\ell}c^\dagger)+ \tilde{G}_{-\ell}(a_{-\ell}^\dagger d + a_{-\ell}d^\dagger),
	\end{align}
	where $a_{\ell(-\ell)}$ denote the annihilation operators for the $|\ell(-\ell)\rangle$ OAM components of the signal mode, and $\tilde{G}_{\ell} = \tilde{G}_{-\ell}$ are the corresponding boosted coupling constants.
	
	In the formalism of second quantization the superposition state $|\psi\rangle$  reads~\cite{Gerry_Knight_2004, PhysRevLett.88.070402, PhysRevA.63.012305} 
	\begin{equation}
		\label{eq:OAM_superposition_state}
		|\psi\rangle = \frac{1}{\sqrt{2}}\left(|1\rangle_{\ell}|0\rangle_{-\ell} + |0\rangle_{\ell}|1\rangle_{-\ell}\right),
	\end{equation}
	where the subscripts denote the two OAM modes. For numerical simulation, it is sufficient to restrict the Hilbert space to the two-mode basis, $|0\rangle_\ell|0\rangle_{-\ell}$, $|0\rangle_\ell|1\rangle_{-\ell}$, $|1\rangle_\ell|0\rangle_{-\ell}$, and $|1\rangle_\ell|1\rangle_{-\ell}$, which can be mapped to the two-qubit computational basis $|\textbf{0}\rangle$, $|\textbf{1}\rangle$, $|\textbf{2}\rangle$, and $|\textbf{3}\rangle$ respectively. The choice of basis is justified by the fact that the atomic sidemodes have negligible excitations ($\bar{n}_{c(d)}\sim0$) at typical BEC temperatures ($\sim 20$ nK), and the state $|\psi\rangle$ contains only a single photon. The loss of photons from the cavity will lead to occupation of the state $|\textbf{0}\rangle \equiv |0\rangle_\ell|0\rangle_{-\ell}$, with some finite probability. In the computational basis, the single photon state is given by $|\psi\rangle \equiv (|\textbf{1}\rangle + |\textbf{2}\rangle)/\sqrt{2}$, which indicates the information encoded in a single qubit~\cite{knill_milburn_linearoptics_2001}. The elements of the initial density matrix $\rho(0)$ and retrieved density matrix $\rho(t\textsubscript{read})$ obtained by solving the relevant Lindblad equation are shown in Figs.~\ref{fig:figure2}(ii)(a) and (b), respectively. The nonzero probability of finding no photon in either OAM mode, resulting from losses due to imperfect state transfer can be seen in Fig.~\ref{fig:figure2}(ii)(b). The overlap between the initial and retrieved states is quantified by the retrieval fidelity defined as $\mathcal{F} = \text{Tr}\left[\sqrt{\rho(0)^{1/2}\rho(t\textsubscript{read})\rho(0)^{1/2}}\right]^2$~\cite{huang_quantum_2015}. The retrieval fidelity as a function of storage time for various $\tilde{G}/\gamma_0$ is presented in Fig.~\ref{fig:figure2}(iii). The classical fidelity benchmark is given by $\mathcal{F}_c = (N+1)/(N+2)$~\cite{PhysRevLett.74.1259, PhysRevLett.94.150503, PhysRevLett.108.190504} where $N$ is the number of qubits. Since, $|\psi\rangle$ is a single OAM qubit state ($N =1)$, $\mathcal{F}_c = 2/3$, indicated by the red dashed line in Fig.~\ref{fig:figure2}(iii). As can be seen from the figure, with a realistic coupling strength of $\tilde{G}/\gamma_{0}=8,$ fidelities of $\sim 88\%$ can be achieved for a storage of about $\sim 0.5$ s which exceeds $\mathcal{F}_c$.
	%
	%
	%
	%
	%
	%
	%
	%
	
	\section{OAM $\pm\ell\hbar$ entangled state}
	
	We now consider the storage of photon pairs entangled in the OAM degree of freedom. These are typically produced using spontaneous parametric down conversion (SPDC)~\cite{mair_entanglement_2001, kwiat_new_1995, doi:10.1098/rsta.2015.0442}. In nondegenerate SPDC, a pump laser carrying OAM $\ell_P\hbar$  and frequency $\omega_P$ is passed through a $\chi^{(2)}$ nonlinear crystal, generating two modes, namely, signal $(\omega_s)$ and idler $(\omega_i)$ with OAMs, $\ell_s\hbar$ and $\ell_i\hbar$, respectively. Due to the conservation of OAM $(\ell_P = \ell_s + \ell_i)$, the signal and idler modes are entangled in the OAM degree of freedom. We consider degenerate SPDC ($\omega_s = \omega_i$) and $\ell_P = 0$ such that if $\ell_s = \ell$ then $\ell_i = -\ell$ and vice versa. Due to linear momentum conservation, the photon pairs are emitted at diametrically opposite points on a cone, illustrated in Fig.~\ref{fig:figure3}(i), where one high-energy photon shown in green is down-converted to produce two lower-energy entangled photons shown in red. By selecting a specific $\ell$ and defining two diametrically opposite paths labeled 1 and 2 on the cone [Fig.~\ref{fig:figure3}(i)], one can postselect a two-photon OAM entangled state $|\psi\rangle_{1,2} = \left(|\ell\rangle_1|-\ell\rangle_2 + |-\ell\rangle_1|\ell\rangle_2\right)/\sqrt{2}$~\cite{mair_entanglement_2001}. 
	
	In the second quantization formalism, the two-photon entangled state is given by 
	\begin{equation}
		\label{eq:OAM_entangled_state}
		|\psi\rangle_{1,2} = \frac{1}{\sqrt{2}} \left(|1,0\rangle_{1}|0, 1\rangle_{2} + |0,1\rangle_{1}|1, 0\rangle_{2}\right),
	\end{equation}
	where, in the notation $|m, n\rangle_i$, $m$($n$) denotes the occupation number of the $|\ell\rangle$ ($|-\ell\rangle$) OAM mode in the path $i$ ($i = 1,2$). This makes $|\psi\rangle_{1,2}$ a four-qubit state. Treating each path as a qubit, the above state can be written in terms of the two-qubit computation basis introduced above as $|\psi\rangle_{1,2} = \left(|\bm{2}\rangle_1|\bm{1}\rangle_2 + |\bm{1}\rangle_1|\bm{2}\rangle_2\right)/\sqrt{2}$. For compact representation of the density matrix, the four-qubit basis $\{|i,j,k, l\rangle\}$ is mapped to $\{|\bm{n}\rangle\}$ so that $|i,j,k,l\rangle = |\bm{p}\rangle_1|\bm{q}\rangle_2 = |\bm{n}\rangle$ with $\bm{n} = 4\bm{p} + \bm{q} - 5$ ($\bm{p},\bm{q} = \bm{1},\bm{2},\bm{3}$). Here, $\bm{n}$ is the binary equivalent of the bit string in $\{|i,j,k,l\rangle\}$.
	
	The photons traveling through paths 1 and 2 are stored in two ring BEC-equipped cavities labeled as 1 and 2, respectively, in Fig.~\ref{fig:figure3}(i) where for pictorial clarity the control fields have been omitted. The retrieved photons from these two cavities are analyzed to verify the preservation of entanglement, following a protocol similar to that of Ref.~\cite{kiesewetter_pulsed_2017}. The Hamiltonian of the composite system is given by
	\begin{equation}
		H\textsubscript{ent} = \sum_{j = 1}^{2}H_j(a_{j,\pm\ell},\tilde{G}_{j,\pm\ell}, c_j, d_j, t),
	\end{equation}
	where $H_{1(2)}$ has the exact form as in Eq.~\eqref{eq:OAM_superposition_Hamiltonian} with the subscript $1(2)$ representing the cavity $1(2)$. The real parts of the initial and retrieved density matrix obtained from solving the relevant Lindblad equation are shown in Figs.~\ref{fig:figure3}(ii)(a) and (b), respectively. In Fig.~\ref{fig:figure3}(ii)(b), nonzero diagonal elements indicate photon loss and the zeros correspond to states forbidden by the conservation of energy, linear momentum, or OAM. To quantify the entanglement $\mathcal{E}_N$ between the photons in paths 1 and 2, we use logarithmic negativity ~\cite{GhobadiPRA2011} (defined in the Appendix~\ref{App-D}) as our measure. The entanglement of the state $|\psi\rangle_{1,2}$ is initially unity for the bipartition between subsystems 1 and 2. It is decreased by the storage process, as reflected in the initial values in Fig.~\ref{fig:figure3}(iii)(a), where $\mathcal{E}_N$ for the retrieved state is plotted versus storage time for different values of $\tilde{G}/\gamma_0$. For $\tilde{G}/\gamma_0 = 8$, $\mathcal{E}_N$ of up to 0.6 can be stored for as long as 0.2 s, with fidelity as high as 0.8, as shown in Figs.~3(iii)(a) and (b), respectively. The red dashed line in Fig.~\ref{fig:figure3}(iii)(b) indicates the average fidelity for teleporting quantum information of dimension $d$ using classical strategies, given by $\mathcal{F}_c = 2/(d+1)$~\cite{PhysRevA.60.1888} which equals $2/5$ for a two-qubit entangled state.
	%
	%
	%
	%
	%
	%
	%
	%
	
	\section{Conclusion}
	
	We have proposed a new platform for the storage and retrieval of quantum states of light encoded in the OAM basis using a ring BEC-cavity system. This platform enables significantly longer storage times ($\sim 0.3$ s) while achieving retrieval fidelities of $\sim90\%$ for the OAM superposition state and $\sim80\%$ for the entangled OAM state, surpassing existing cold-atom-based quantum memory protocols for OAM-encoded qubits, which are typically limited to hundreds of microseconds. Further, our scheme allows all atoms to remain in the same interal ground state, and offers $\sim 10^{4}$ motional states for storage and $\sim 10^{5}$ transitions for reading and writing. Finally, due to the equivalence of the system Hamiltonian with that of standard optomechanical systems, our model naturally supports OAM transduction~\cite{Kaviani:20}, wavelength conversion~\cite{PhysRevA.82.053806}, and non-destructive readout~\cite{kumar_cavity_2021}, broadening its potential for integration into quantum information networks.
	%
	%
	%
	\begin{acknowledgments}
		M.B. thanks the Air Force Office of Scientific Research (AFOSR) (FA9550-23-1-0259) for support. N.D. acknowledges Sarmad Maqsood for assistance with the diagrams and Jhair Gallardo for setting up the computational environment.  
		R.G. acknowledges financial support from CSIR-HRDG, India. P.K.  acknowledges the financial support provided by the Max Planck Society. H.S.D. acknowledges funding from SERB/ANRF-DST, India (CRG/2021/008918) and IRCC, IIT Bombay (RD/0521-IRCCSH0-001).
		We are grateful for stimulating conversations with T. Chakraborty, T.N. Dey, and A. Sen(De). All simulation results in this work were obtained using the open source Python library QuTiP~\cite{qutip}.
	\end{acknowledgments}
	
\onecolumngrid
\appendix
	
	\section{System Hamiltonian\label{App-A}}
	This Appendix details: (i) the full system Hamiltonian [Eq.~\eqref{eq:ring_BEC_Hamiltonian_1}] and its linearization [Eq.~\eqref{eq:single_mode_linearized_Hamiltonian}], (ii) a protocol to generate single photons in coherent superpositions of $\pm \ell\hbar$ orbital angular momentum (OAM) states, and (iii) comprehensive fidelity analysis with logarithmic negativity formalism for entanglement quantification.

	Here we establish the full Hamiltonian for a ring Bose-Einstein condensate (BEC) [sodium ($^{23}$Na)] confined within a Fabry-Perot cavity \cite{kumar_cavity_2021}, driven by both control and signal laser fields. Within the toroidal trap, atoms of mass $m$ in the $N$-atom condensate experience a potential
	\begin{equation}
		\label{eq:harmonic_potential}
		U(\rho,z) = \frac{1}{2}m\omega_\rho\left(\rho - R\right)^2 + \frac{1}{2}m\omega_z z^2,
	\end{equation}
	where blue$R$ is the ring radius, and $\omega_\rho$, $\omega_z$ are the harmonic trapping frequencies along the radial ($\rho$) and axial directions ($z$), respectively. This potential results in free atomic motion along the ring's azimuthal ($\phi$) direction.\\
	
    \subsection{Many-Body Hamiltonian}
	We first derive the single-atom Hamiltonian before extending to the full $N$-atom ring BEC system. The cavity in our system is driven by two optical fields: a strong control laser (frequency $\omega_L$, pump rate $\varepsilon_c$) and a weak signal laser (frequency $\omega_p$, pump rate $\varepsilon_p$). Blue-detuned from the atomic transition, both fields are prepared in coherent superpositions of Laguerre-Gaussian (LG) modes~\cite{Yao:11} carrying $\pm \ell\hbar$ OAM. These superpositions generate a circular optical lattice around the cavity axis that interacts with the ring BEC, which exhibits rotational motion characterized by the winding number $L_p$. Employing a two-level atom approximation with dispersive light-matter interactions \cite{Gerry_Knight_2004}, the single-particle Hamiltonian for azimuthal motion (along $\phi$-direction) in the control field's rotating frame is:
	\begin{align}
		H_{\text{single}} = 
		\underbrace{-\frac{\hbar^{2}}{2I}\frac{\partial^{2}}{\partial\phi^{2}}}_{\substack{\text{Rotational} \\ \text{kinetic energy}}} 
		- \underbrace{\hbar\Delta_{0}a^{\dagger}a}_{\substack{\text{Detuned} \\ \text{cavity energy}}} 
		+ \underbrace{\hbar U_{0}\cos^{2}(\ell\phi)a^{\dagger}a}_{\substack{\text{Optical lattice} \\ \text{potential}}} + \underbrace{i\hbar(\varepsilon_c a^\dagger - \varepsilon_c^* a)}_{\substack{\text{Control field} \\ \text{drive}}} 
		+ \underbrace{i\hbar(\varepsilon_p a^\dagger e^{-i\delta t} - \varepsilon_p^* a e^{i\delta t})}_{\substack{\text{Signal field} \\ \text{drive}}} \;,
		\label{eq:Single_Particle_Hamiltonian}
	\end{align}
	where $I=mR^{2}$ is the moment of inertia of a single atom about the cavity axis; $a$ is the bosonic annihilation operator for the intra-cavity field satisfying $[a, a^\dagger] = 1$; $\Delta_0 = \omega_L - \omega_0$ denotes the control laser detuning from cavity resonance; $U_0 = g_a^2 / \Delta_a$ represents the atom-lattice coupling strength, with $g_a$ the single-atom-photon coupling and $\Delta_a$ the atomic detuning;  $\varepsilon_{c(p)}=\sqrt{\frac{P_{in}^{c(p)}\gamma_{0}}{\hbar\omega_{0}}}$ where $P_{in}^{c(s)}$ is the power of control (signal) field; and $\delta=\omega_{p}-\omega_{c}$ is the two-photon detuning.

	The many-body Hamiltonian describing azimuthal dynamics of atoms in a ring BEC, incorporating two-body interactions, is:
	\begin{align}
		H = 
		\underbrace{\int_{0}^{2\pi} d\phi \, \Psi^{\dagger}(\phi) H_{\mathrm{single}} \Psi(\phi)}_{\substack{\text{Single-particle contributions}}}
		+ 
		\underbrace{\frac{g}{2} \int_{0}^{2\pi} d\phi \, \Psi^{\dagger}(\phi) \Psi^{\dagger}(\phi) \Psi(\phi) \Psi(\phi)}_{\substack{\text{Two-body atomic interactions}}}\;,\label{eq:ManyBody_Hamiltonian}
	\end{align}
	where $\Psi(\phi)$ is the bosonic atomic field operator satisfying $[\Psi(\phi),\Psi^{\dagger}(\phi')]=\delta(\phi-\phi')$, and $g=2\hbar\omega_{\rho}a_{\mathrm{Na}}/R$ represents the two-body interaction strength, with $a_{\mathrm{Na}}$ being $s$-wave scattering length of sodium. Substituting Eq. (\ref{eq:Single_Particle_Hamiltonian}) in Eq. (\ref{eq:ManyBody_Hamiltonian}) yields the full many-body Hamiltonian:
	\begin{align}
		H &=\int_{0}^{2\pi} d\phi \, \Psi^{\dagger}(\phi)\Bigg[-\frac{\hbar^{2}}{2I}\frac{\partial^{2}}{\partial\phi^{2}}+\hbar U_{0}\cos^{2}(\ell\phi)a^{\dagger}a\Bigg]\Psi(\phi)
		-\hbar\Delta_{0}a^{\dagger}a+i\hbar(\varepsilon_c a^\dagger - \varepsilon_c^* a)+i\hbar(\varepsilon_p a^\dagger e^{-i\delta t} - \varepsilon_p^* a e^{i\delta t})\nonumber\\
		&+\frac{g}{2} \int_{0}^{2\pi} d\phi \, \Psi^{\dagger}(\phi) \Psi^{\dagger}(\phi) \Psi(\phi) \Psi(\phi)\;.\label{eq:ManyBody_FullHamiltonian}
	\end{align}

    \subsection{Canonical Optomechanical Hamiltonian}
	The weak optical lattice induces coherent first-order Bragg scattering in a fraction of condensate atoms, transferring them from original rotational states with winding number $L_{p}$ to $L_{p}\pm 2\ell$. This physical process motivates us to express the atomic field operator as
	\begin{align}
		\Psi(\phi)=\frac{1}{\sqrt{2\pi}}\Big[e^{iL_{p}\phi}+e^{i(L_{p}+2\ell)\phi}+e^{i(L_{p}-2\ell)\phi}\Big]\;,\label{eq:Ansatz}
	\end{align}
	where the atomic operators satisfy the canonical commutation relations $[c_i, c_j^\dagger] = \delta_{ij}$ for $i,j \in {p, +, -}$, and the total atom number operator $c_p^\dagger c_p + c_+^\dagger c_+ + c_-^\dagger c_- = N$ gives the condensate population. Substituting the atomic field ansatz into Eq. (\ref{eq:ManyBody_FullHamiltonian}) under the side-mode depletion approximation ($\langle c_\pm^\dagger c_\pm \rangle \ll \langle c_p^\dagger c_p \rangle$), we obtain optomechanical-type Hamiltonian [Eq. (1) in main text]: 
	\begin{align}
		H &= \underbrace{\hbar\omega_c c^\dagger c + \hbar\omega_d d^\dagger d}_{\text{Atomic side-modes}} 
		+ \underbrace{\hbar G( X_c +  X_d) a^\dagger a}_{\text{Optomechanical coupling}} 
		- \underbrace{\hbar\tilde{\Delta} a^\dagger a}_{\text{Detuned cavity}} \nonumber \\
		& + \underbrace{i\hbar\left(\varepsilon_c a^\dagger - \varepsilon^*_c a\right) + i\hbar\left(\varepsilon_p a^\dagger e^{-i\delta t} - \varepsilon^*_p ae^{i\delta t}\right)}_{\text{Control/probe drives}} 
		+ \underbrace{4\hbar\tilde{g}N(c^\dagger c + d^\dagger d) + 2\hbar\tilde{g}N(c d + c^\dagger d^\dagger)}_{\text{Atomic interactions}}. 
		\label{eq:ring_BEC_Hamiltonian}
	\end{align}
	
	The Hamiltonian components represent:
	\begin{enumerate}
		\item \textbf{Atomic side-mode excitations:} The $c$ and $d$ operators describe quantum fluctuations of rotational excitations, defined via 
		\begin{align}
			c = \frac{c_p^* c_+}{\sqrt{N}},\quad d = \frac{c_p^* c_-}{\sqrt{N}}\;,\label{eq:SideMode_Operators}
		\end{align}
		where $c_p$ represents the complex amplitude of the macroscopically occupied primary condensate mode. Their frequencies 
		\begin{align}
			\omega_c = \frac{\hbar(L_p + 2\ell)^2}{2I}, \quad \omega_d = \frac{\hbar(L_p - 2\ell)^2}{2I}\;,\label{eq:SideMode_Frequencies}
		\end{align}
		reflect quadratic angular momentum dependence.
		\item \textbf{Optomechanical coupling:} The $G(X_c + X_d)a^\dagger a$ term  mediates light-atomic sidemode interactions with strength $G = U_0\sqrt{N/8}$, linking cavity photons to atomic position quadratures, $X_j \equiv (j + j^\dagger)/\sqrt{2}$ ($j=c,d$).
		\item \textbf{Detuned cavity energy:} The $-\hbar\tilde{\Delta}a^{\dagger}a$ term represents free energy of the cavity with a modified detuning  $\tilde{\Delta} = \Delta_0 + U_0 N / 2$.
		\item \textbf{Control/signal laser drives:} The fifth and sixth terms in Eq.~\eqref{eq:ring_BEC_Hamiltonian} represent cavity drive with a strong control and a weak signal laser fields, respectively.
		\item \textbf{Atomic interactions:}
		The $\tilde{g}$-dependent terms ($\tilde{g}=g/4\pi\hbar$) include mean-field energy shifts and anomalous pairing $2\hbar\tilde{g}N(c d + c^\dagger d^\dagger)$ enabling simultaneous creation or destruction of pairs of distinct atomic excitations.
	\end{enumerate}
	
    \subsection{Linearized Hamiltonian for Quantum Memory}
	To derive the linearized Hamiltonian in Eq. (2) of the main text, we assume the weak signal limit ($\varepsilon_p \ll \varepsilon_c$) and express the intra-cavity field and Bragg-scattered sidemodes as follows:
	\begin{equation}
		\label{eq:linearlization_approximation_2}
		a = \alpha +  \tilde{a}e^{-i\delta t},\;  c = \beta_c +  \tilde{c}e^{-i\delta t},\text{ and }  d = \beta_d +  \tilde{d}e^{-i\delta t}\;,
	\end{equation}
	where, $\alpha = \langle a\rangle$ is the mean value of the intra-cavity field in absence of the signal field, $\beta_{c(d)} = \langle c(d)\rangle$ is the mean value of the sidemode $c(d)$, and $ \tilde{a}$ is quantum field operator for the signal mode. To linearize the Hamiltonian in Eq.~\eqref{eq:ring_BEC_Hamiltonian}, we employ the decomposition from Eq.~\eqref{eq:linearlization_approximation_2} to express
	
\begin{align}
H &=\hbar\sum_{j=c,d}\omega_{j}\Bigg[j^{\dagger}j+\Big(\beta_{j}j^{\dagger}e^{i\delta t}+\beta_{j}^{\ast}je^{-i\delta t}\Big)\Bigg]+\frac{\hbar G}{\sqrt{2}}\sum_{j=c,d}\Bigg[\Big(\beta_{j}+\beta_{j}^{\ast}\Big)\Big(\alpha \tilde{a}^{\dagger}e^{i\delta t}+\alpha^{\ast} \tilde{a}e^{-i\delta t}\Big)+\Big(\beta_{j}+\beta_{j}^{\ast}\Big)a^{\dagger}\tilde{a}\nonumber\\
&\quad + \Big(j^{\dagger}e^{i\delta t}+je^{-i\delta t}\Big)|\alpha|^{2}+\Big(\alpha j^{\dagger}\tilde{a}e^{2i\delta t}+\alpha^{\ast}j\tilde{a}e^{-2i\delta t}\Big)+\Big(\alpha j \tilde{a}^{\dagger}+\alpha^{\ast} j^{\dagger}\tilde{a}\Big)\Bigg]-\hbar\tilde{\Delta}\Bigg[\tilde{a}^{\dagger}\tilde{a}+\Big(\alpha \tilde{a}^{\dagger}e^{i\delta t}+\alpha^{\ast}\tilde{a}e^{-i\delta t}\Big)\Bigg]\nonumber\\
&\quad +i\hbar\varepsilon_{c}\Bigg[\tilde{a}^{\dagger}e^{i\delta t}-\tilde{a}e^{i\delta t}\Bigg]+4\hbar\tilde{g}N\sum_{j=c,d}\Bigg[j^{\dagger}j+\Big(\beta_{j}j^{\dagger}e^{i\delta t}+\beta_{j}^{\ast}j e^{-i\delta t}\Big)\Bigg]\nonumber\\
&\quad +2\hbar\tilde{g}N\Bigg[\Big(\beta_c^{\ast}d^{\dagger}+\beta_d^{\ast}c^{\dagger}\Big)e^{i\delta t}+\Big(\beta_c d+\beta_d c\Big)e^{-i\delta t}+\Big(c^{\dagger}d^{\dagger}e^{2i\delta t}+cd e^{-2i\delta t}\Big)\Bigg]+\mathcal{C}\;,\label{eq:linearized_Hamiltonian_1}
\end{align}
In deriving the above Hamiltonian, we have neglected cubic fluctuation terms such as $ja^{\dagger}\tilde{a}$ and $j^{\dagger} a^{\dagger}\tilde{a}$ ($j=c,d$) and also assumed real pumping rates for control and signal. The symbol  $\mathcal{C}$ denotes a constant term given by:
\begin{align}
\mathcal{C}&= -\hbar\tilde{\Delta}|\alpha|^{2}+\hbar\sum_{j=c,d}|\beta_{j}|^{2}\omega_{j}+\frac{\hbar G}{\sqrt{2}}\sum_{j=c,d}(\beta_{j}+\beta_{j}^{\ast})|\alpha|^{2}+i\hbar\varepsilon_{c}(\alpha^{\ast}-\alpha) +4\hbar\tilde{g}N\sum_{j=c,d}|\beta_{j}|^{2}+2\hbar\tilde{g}N(\beta_c \beta_d +\beta_c^{\ast}\beta_d^{\ast})\;,\label{eq:constant_term}
\end{align}
	After removing the constant term in Eq.~\eqref{eq:constant_term} via energy shift, we redefine Eq. (\ref{eq:linearized_Hamiltonian_1}) with boosted coupling $\tilde{G}=G\alpha/\sqrt{2}$ and modified detuning $\Delta^\prime = \tilde{\Delta} - \frac{G}{\sqrt{2}} \sum_{j=c,d} (\beta_j + \beta_j^*)$ as:
	\begin{align}
		H &=-\hbar\Delta^\prime a^{\dagger}\tilde{a}+\hbar\sum_{j=c,d}\omega_{j}j^{\dagger}j+\hbar\sum_{j=c,d}\Bigg[\tilde{G} j^{\dagger}\tilde{a}e^{2i\delta t}+\tilde{G}^{\ast}j\tilde{a}e^{-2i\delta t}\Bigg]+\hbar\sum_{j=c,d}\Bigg[\tilde{G} j\tilde{a}^{\dagger}+\tilde{G}^{\ast}j^{\dagger}\tilde{a}\Bigg]\nonumber\\
		&\quad +i\hbar \tilde{a}^{\dagger}\Bigg\{i\tilde{\Delta}\alpha-i\tilde{G}\sum_{j=c,d}(\beta_{j}+\beta_{j}^{\ast})+\varepsilon_{c}\Bigg\}e^{i\delta t}-i\hbar \tilde{a}\Bigg\{-i\tilde{\Delta}\alpha^{\ast}+i\tilde{G}^{\ast}\sum_{j=c,d}(\beta_{j}+\beta_{j}^{\ast})+\varepsilon_{c}\Bigg\}e^{-i\delta t}\nonumber\\
		&\quad+i\hbar\Bigg\{c^{\dagger}\Big(-i\omega_{c}\beta_c-i\tilde{G}\alpha^{\ast}-4i\tilde{g}N\beta_c-2i\tilde{g}N\beta_d^{\ast}\Big)\Bigg\}e^{i\delta t}-i\hbar\Bigg\{c\Big(i\omega_{c}\beta_c^{\ast}+i\tilde{G}^{\ast}\alpha+4i\tilde{g}N\beta_c^{\ast}+2i\tilde{g}N\beta_d\Big)\Bigg\}e^{-i\delta t}\nonumber\\
		&\quad+i\hbar\Bigg\{d^{\dagger}\Big(-i\omega_{d}\beta_d-i\tilde{G}\alpha^{\ast}-4i\tilde{g}N\beta_d-2i\tilde{g}N\beta_c^{\ast}\Big)\Bigg\}e^{i\delta t}-i\hbar\Bigg\{d\Big(i\omega_{d}\beta_d^{\ast}+i\tilde{G}^{\ast}\alpha+4i\tilde{g}N\beta_d^{\ast}+2i\tilde{g}N\beta_c\Big)\Bigg\}e^{-i\delta t}\nonumber\\
		&\quad+4\hbar\tilde{g}N\sum_{j=c,d}j^{\dagger}j+2\hbar\tilde{g}N\Bigg[c^{\dagger}d^{\dagger} e^{2i\delta t}+cd e^{-2i\delta t}\Bigg]\;,\label{eq:linear_Hamiltonian_2}
	\end{align}
	The Heisenberg equations of motion for the expectation values $\langle a\rangle = \alpha$ and $\langle c(d)\rangle = \beta_{c(d)}$ in absence of the probe field are
	\begin{align*}
		\partial_t\alpha &= -\left(\frac{\gamma_0}{2} -i\Delta^\prime\right)\alpha -i\tilde{G}\sum_{j=c,d}(\beta_{j}+\beta_{j}^{\ast}) + \varepsilon_c,\\
		\partial_t\beta_c &= -\left(\frac{\gamma_m}{2} + i\omega_c\right)\beta_c  -i \tilde{G}\alpha^{\ast}-4i\tilde{g}N\beta_c-2i\tilde{g}N\beta_d^{\ast}\;,\\
		\partial_t\beta_d &= -\left(\frac{\gamma_m}{2} + i\omega_d\right)\beta_d  -i \tilde{G}\alpha^{\ast}-4i\tilde{g}N\beta_d-2i\tilde{g}N\beta_c^{\ast}\;.
	\end{align*}
	The terms inside the curly brackets of Eq.~\eqref{eq:linear_Hamiltonian_2} are simply the right-hand side of the above equations, except the decay terms $\gamma_0/2$ and $\gamma_m/2$. The right-hand side of the above equations go to zero in steady state, which is reached very early. Therefore, if we are only interested in the fluctuations around the average, then the curly bracketed terms can be put to zero. The linearized Hamiltonian can then be written as
	\begin{align}
		H &=-\hbar\Delta^\prime \tilde{a}^{\dagger}\tilde{a}+\hbar\sum_{j=c,d}\omega_{j}j^{\dagger}j+\hbar\sum_{j=c,d}\Bigg[\tilde{G} j^{\dagger}\tilde{a}e^{2i\delta t}+\tilde{G}^{\ast}j\tilde{a}e^{-2i\delta t}\Bigg]+\hbar\sum_{j=c,d}\Bigg[\tilde{G} j\tilde{a}^{\dagger}+\tilde{G}^{\ast}j^{\dagger}\tilde{a}\Bigg]\nonumber\\
		&\quad+4\hbar\tilde{g}N\sum_{j=c,d}j^{\dagger}j+2\hbar\tilde{g}N\Bigg[c^{\dagger}d^{\dagger} e^{2i\delta t}+cd e^{-2i\delta t}\Bigg]\;,\label{eq:linear_Hamiltonian_3}
	\end{align}
	After neglecting rapidly oscillating terms, the linearized Hamiltonian above takes the following form:
	\begin{align}
		H &=-\hbar\Delta^\prime\tilde{a}^{\dagger}\tilde{a}+\hbar\sum_{j=c,d}(\omega_{j}+4\tilde{g}N)j^{\dagger}j+\hbar\sum_{j=c,d}\Bigg[\tilde{G} j\tilde{a}^{\dagger}+\tilde{G}^{\ast}j^{\dagger}\tilde{a}\Bigg]\;.\label{eq:linear_Hamiltonian_final}
	\end{align}
	\subsubsection{Resolved sideband limit}
	Setting $\delta = \omega_c$ and working in the resolved sideband limit ($\gamma_0 \ll \omega_c$) with $\omega_c \gg \omega_d$ (making mode $d$ off-resonant), we simplify Eq.~\eqref{eq:linear_Hamiltonian_final} by choosing $\Delta^\prime = \tilde{\Delta}$ for typical parameters and removing the tilde above the operators, yielding:
	\begin{equation}
		\label{eq:linear_Hamiltonian_modeCwithCollisions}
		H = -\hbar \tilde{\Delta}  a^\dagger a + \hbar(\omega_c+4\tilde{g}N)  c^\dagger c + \hbar \left( \tilde{G}  a^\dagger c + \tilde{G}^*  a c^\dagger \right). 
	\end{equation}
	Ignoring the atomic interactions and applying the unitary transformation $U = \exp[-i\delta t  a^\dagger a] \otimes \exp[-i\delta t  c^\dagger c]$ with $\tilde{\Delta} = -\delta$ eliminates the free evolution terms, resulting in the quantum state transfer Hamiltonian:
	\begin{equation}
		\label{eq:linear_Hamiltonian}
		H = \hbar \left( \tilde{G}  a^\dagger c + \tilde{G}^*  a c^\dagger \right),
	\end{equation}
	which corresponds to Eq. (2) in the main text. This beam-splitter interaction facilitates quantum state transfer from optical mode $a$ to atomic mode $c$.\\
	For the storage and retrieval of OAM supersposition state, two control fields with distinct frequencies are required. Following the linearization procedure as above, the Hamiltonian as given in Eq. (4) of the main text can be derived. In this case, linearization of the atomic interaction term gives,
	\begin{equation}
		4\hbar \tilde{g} N \left( c^\dagger c + d^\dagger d \right)
		+ 2\hbar \tilde{g} N \left( c d e^{-i(\delta_1 + \delta_2) t} + c^\dagger d^\dagger e^{i (\delta_1 + \delta_2) t} \right),
	\end{equation}
	where, $\delta_1 = \omega_c$ and $\delta_2 = \omega_d$.
	
	
	\subsection{Parameter Space Boundaries}
	When evaluating $\tilde{G}$, the following parameter constraints must be satisfied:
	\begin{itemize}
		\item Quasi-1D dynamics in the ring geometry must be maintained, which imposes an upper bound on atom number:
		\begin{align}
			\label{eq:constraint_1}
			N<\frac{4R}{3a_{\mathrm{Na}}}\sqrt{\frac{\pi\omega_{\rho}}{\omega_{z}}}\;.
		\end{align}
		\item The Bogoliubov-dressed side mode frequencies $\omega_{c,d}' = \sqrt{\omega_{c,d} ( \omega_{c,d} + 4\tilde{g}N )}$ reduce to the bare frequencies $\omega_{c,d}$ under the condition:
		\begin{align}
			\omega_{c,d} \gg 4\tilde{g}N\;,\label{eq:constraint_2}
		\end{align}
		which we impose to enable the approximation $\omega_{c,d}' \approx \omega_{c,d}$. 
		\item The optical lattice potential must remain weaker than the rotational BEC's chemical potential to preserve condensate coherence, requiring:
		\begin{align}
			\hbar U_0|\alpha|^2 \ll \frac{\hbar^2 L_p^2}{2mR^2} + 2\hbar\tilde{g}N\;.\label{eq:constraint_3}
		\end{align}
	\end{itemize}
	
	\subsection{Experimentally Feasible Parameters}
	The relevant experimentally feasible parameters~\cite{PhysRevA.107.013525} that satisfy the constraints given in Eqs.~\eqref{eq:constraint_1}, \eqref{eq:constraint_2}, and \eqref{eq:constraint_3} are $m = 23$ amu, $a\textsubscript{Na} = 0.1$ nm, $g_a/2\pi = 0.36$ MHz, $\Delta_a = 300\Gamma$ where $\Gamma/2\pi = 9.8$ MHz is the decay rate of Sodium D-line, $\omega_\rho/2\pi = \omega_z/2\pi = 840$ Hz, $R = 10$ $\mu$m, $\gamma_0/2\pi = 1$ kHz, $\gamma_m/2\pi \sim 1.7\times 10^{-5}\gamma_0$, and $\ell = 130$. Achieving $\tilde{G}/\gamma_0 \sim 4$ as mentioned in the main manuscript requires control power, $P_{in}^{c} \sim 8.6\times10^{-7}$ mW, $\mathcal{P}^s_{in} = 8.6\times 10^{-9}$ mW, $L_p = 20$, and $N = 2\times10^4$. For, $\tilde{G}/\gamma_0 \sim 8$, everything else remains same except, $P_{in}^{c} \sim 1.45\times10^{-6}$ mW, $L_p = 25$, and $N = 8\times10^4$.
	
	
    \section{Superposition generation\label{App-B}}
	To generate a single photon in a coherent superposition of $\pm\hbar\ell$ OAM states as given in Eq. (5) of the main text, a Mach-Zehnder interferometer setup is used as illustrated in Fig.~\ref{fig:figure_a1}:
	\begin{figure}[h]
		\centering
		\includegraphics[width=\linewidth]{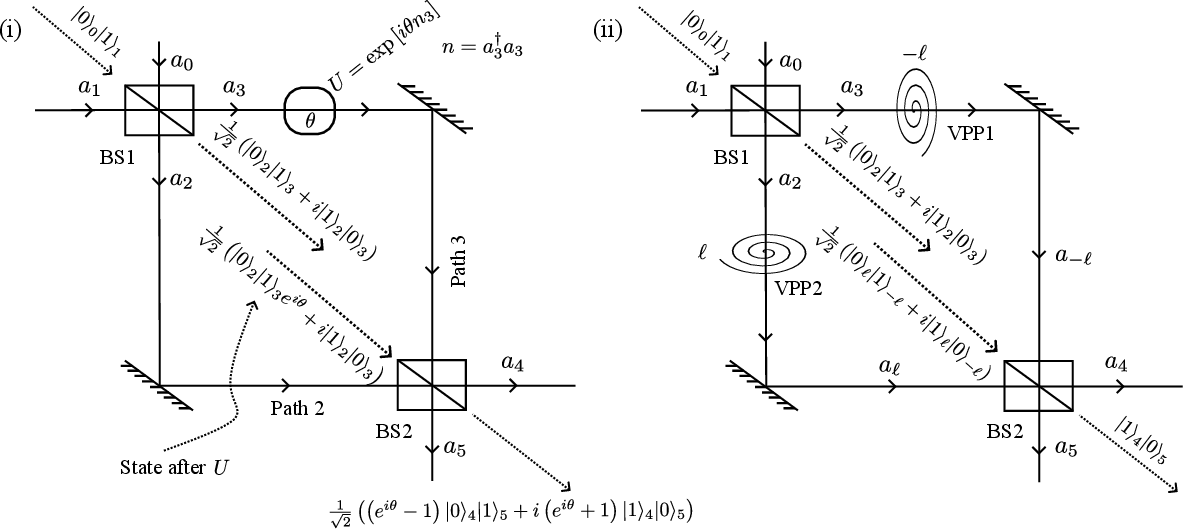}
		\caption{\label{fig:figure_a1}(i) Schematic diagram of Mach-Zehnder interferometer. The different bosonic modes representing the paths taken by the photon are denoted by $a_0 -  a_5$, $U$ is a transformation corresponding to the phase element placed in the path of mode $ a_2$. The beam splitters are denoted by BS. (ii) Schematic diagram for generation of a single photon in a coherent superposition of $|\pm\ell\rangle$ OAM state. The vortex phase plates to generate a photon with $+(-)\ell\hbar$ OAM is denoted by VPP1(2). The modes, $ a_2\equiv a_{\ell}$ and $ a_3\equiv a_{-\ell}$ denote the annihilation operators for OAM modes $|+\ell\rangle$ and $-|\ell\rangle$, respectively.}
	\end{figure}
	Assuming that the reflected beam suffers a $\pi/2$ phase shift, the input and output modes for beam splitter 1 (BS1) and BS2 are related according to
	\begin{equation}
		a_2 = \frac{1}{\sqrt{2}}\left( a_0 + i a_1\right), \quad  a_3 = \frac{1}{\sqrt{2}}\left( a_1 + i a_0\right),\quad\text{and}\quad  a_4 = \frac{1}{\sqrt{2}}\left( a_2 + i a_3\right), \quad  a_5 = \frac{1}{\sqrt{2}}\left( a_3 + i a_2\right),
	\end{equation}
	respectively. From the above equations, we have
	\begin{equation}
		\label{eq:a0_and_a1}
		a_0 = \frac{1}{\sqrt{2}}\left( a_2 - i a_3\right), \quad  a_1 = \frac{1}{\sqrt{2}}\left( a_3 - i a_2\right),\quad  a_2 = \frac{1}{\sqrt{2}}\left( a_4 - i a_5\right), \quad  a_3 = \frac{1}{\sqrt{2}}\left( a_5 - i a_4\right),
	\end{equation}
	In Fig.~\ref{fig:figure_a1}, the input state of beam splitter 1 (BS1) is $|0\rangle_0|1\rangle_1$ which can be written as:
	\begin{equation}
		|0\rangle_0|1\rangle_1|0\rangle_2|0\rangle_3 =  a^\dagger_1|0\rangle_0|0\rangle_1|0\rangle_2|0\rangle_3 = \frac{1}{\sqrt{2}}\left( a_2 - i a_3\right)^\dagger|0\rangle_0|0\rangle_1|0\rangle_2|0\rangle_3= \frac{1}{\sqrt{2}}|0\rangle_0|0\rangle_1\left(|0\rangle_2|1\rangle_3 + i|1\rangle_2|0\rangle_3\right),
	\end{equation}
	For brevity, the above equation is written as
	\begin{equation}
		\underbrace{|0\rangle_0|1\rangle_1}_\text{in}\xrightarrow{BS1}\underbrace{\frac{1}{\sqrt{2}}\left(|0\rangle_2|1\rangle_3 + i|1\rangle_2|0\rangle_3\right)}_\text{out},
	\end{equation}
	i.e., one photon enters BS1 through mode $ a_1$. After the action BS1, the single photon can occupy the mode $ a_2$ or $ a_3$.

	In Fig.~\ref{fig:figure_a1}(i), a phase element is added along path 3 output state of BS1 as
	\begin{equation}
		\label{eq:output_of_BS1}
		\frac{1}{\sqrt{2}}\left(|0\rangle_2|1\rangle_3 + i|1\rangle_2|0\rangle_3\right)\xrightarrow{\hat{U}(\theta)}	\frac{1}{\sqrt{2}}\left(|0\rangle_2|1\rangle_3e^{i\theta} + i|1\rangle_2|0\rangle_3\right).
	\end{equation}
	The state in Eq.~\eqref{eq:output_of_BS1} is an input to BS2 whose output state now becomes
	\begin{equation}
		\frac{1}{\sqrt{2}}\left(|0\rangle_2|1\rangle_3e^{i\theta} + i|1\rangle_2|0\rangle_3\right)\xrightarrow{BS2}\frac{1}{2}\left[\left(e^{i\theta} - 1\right)|0\rangle_4|1\rangle_5 + \left(e^{i\theta} + 1\right)|1\rangle_4|0\rangle_5\right].
	\end{equation}
	Therefore, in the absence of the phase element, the output state of BS2 is: 
	\begin{equation}
		\label{eq:output_of_BS2}
		|\psi\rangle\textsubscript{BS2(out)} = |1\rangle_4|0\rangle_5
	\end{equation}
	i.e., there will always be one photon in mode $ a_4$ and zero photons in mode $ a_5$.
	
	For generating a photon that is in a coherent superposition of $\ell\hbar$ and $-\ell\hbar$ OAM modes, a vortex plate of $\ell$ OAM index is added in the path of mode $ a_2$ and another vortex phase plate of $-\ell$ OAM index is added in the path of mode $ a_3$ as shown in Fig.~\ref{fig:figure_a1}(ii). Therefore, the mode $ a_{2(3)}$ can be equivalently written as $ a_{\ell(-\ell)}$ and the modes $ a_4$ and $ a_5$ as
	\begin{equation}
		\label{eq:al_and_a_minus_l}
		a_4 = \frac{1}{\sqrt{2}}\left( a_{\ell} + i a_{-\ell}\right), \quad  a_5 = \frac{1}{\sqrt{2}}\left( a_{-\ell} + i a_{\ell}\right)
	\end{equation}
	Therefore, the input state of BS2 can be equivalently written as:
	\begin{equation}
		|\psi\rangle\textsubscript{BS2(in)}\equiv\frac{1}{\sqrt{2}}\left(|0\rangle_\ell|1\rangle_{-\ell} + i|1\rangle_\ell|0\rangle_{-\ell}\right).
	\end{equation}
	If there are no phase elements (Fig.~\ref{fig:figure_a1}), the output of BS2 is always
	\begin{equation}
		|\psi\rangle\textsubscript{BS2(out)} = |1\rangle_4|0\rangle_5.\quad\left( \text{From, Eq.~\ref{eq:output_of_BS2}}\right)
	\end{equation}
	The single photon coming out of mode $ a_4$ can come from transmission of mode $ a_2\equiv a_{\ell}$ or reflection of mode $ a_3\equiv a_{-\ell}$ i.e., this single photon in mode $ a_4$ is a superposition of modes $ a_{\ell}$ and $ a_{-\ell}$ as evident from Eq.~\eqref{eq:al_and_a_minus_l}.
	%
	%
	%
	\section{Fidelity analysis\label{App-C}}
\begin{figure}[h]
		\centering
		\includegraphics[width=0.5\linewidth]{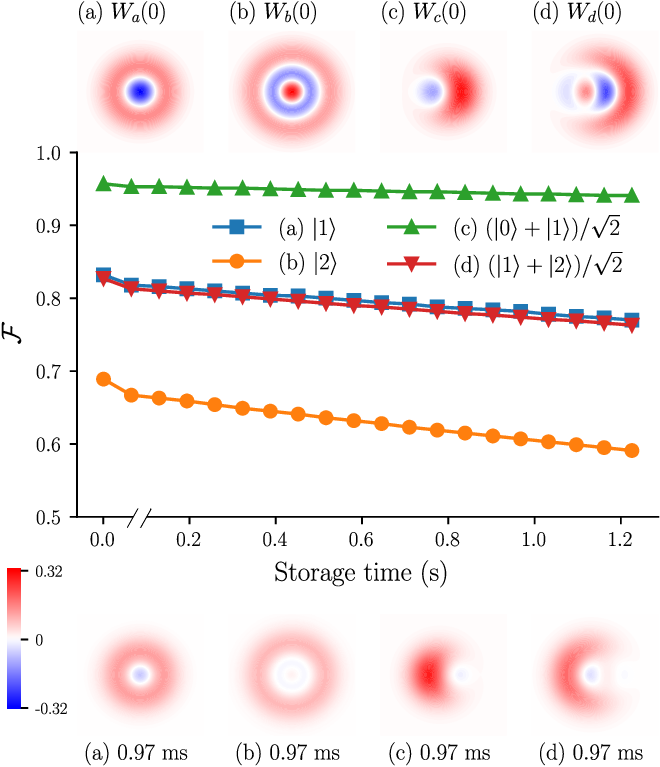}
		\caption{\label{fig:figure_a2} Fidelity versus storage time for (a) $|1\rangle_\ell$, (b) $|2\rangle_\ell$, (c) $(|0\rangle_\ell + |1\rangle_\ell)/\sqrt{2}$, and (d) $(|1\rangle_\ell + |2\rangle_\ell)/\sqrt{2}$. The top row shows the intial Wigner functions for the above states, respectively. Bottom row shows the corresponding retrieved Wigner functions at $\sim 1$ ms. Parameters used: Control power$\sim8.6\times10^{-7}$ mW, $L_p = 20$, $\ell = 130$, $N = 2\times10^4$, $\gamma_0/2\pi = 1$ KHz, $\gamma_m/2\pi \sim 1.7\times 10^{-5}\gamma_0$, $\tilde{G}/2\pi = 4.1\gamma_0$, $t\textsubscript{off} = \pi/2\tilde{G} \sim 61$ $\mu$s.}
	\end{figure}

	%
	For completeness, the variation of fidelity versus storage time for some Fock states and their number basis superpositions in the OAM mode $|\ell\rangle$ is shown in Fig.~\ref{fig:figure_a2}. The definition of fidelity used is
	\begin{equation}
		\mathcal{F} = \left[\text{Tr}\left(\sqrt{\sqrt{\rho(0)}\rho(t\textsubscript{read})\sqrt{\rho(0)}}\right)\right]^2,
	\end{equation}
	where, $\rho(0)$ is the initial density matrix of the system and $\rho(t\textsubscript{read})$ is the retrieved density matrix.
	As can be seen from Fig.~\ref{fig:figure_a2}, Wigner function's negativity is retained for fock states, $|1\rangle_\ell$, $|2\rangle_\ell$, $\left(|0\rangle_\ell + |1\rangle_\ell\right)/\sqrt{2}$, and $\left(|0\rangle_\ell + |1\rangle_\ell\right)/\sqrt{2}$ after a storage of about $1$ ms which is much larger than what is currently achievable, typically in microseconds~\cite{PhysRevLett.122.210501}.

\begin{figure}[h]
		\centering
		\includegraphics[width=\linewidth]{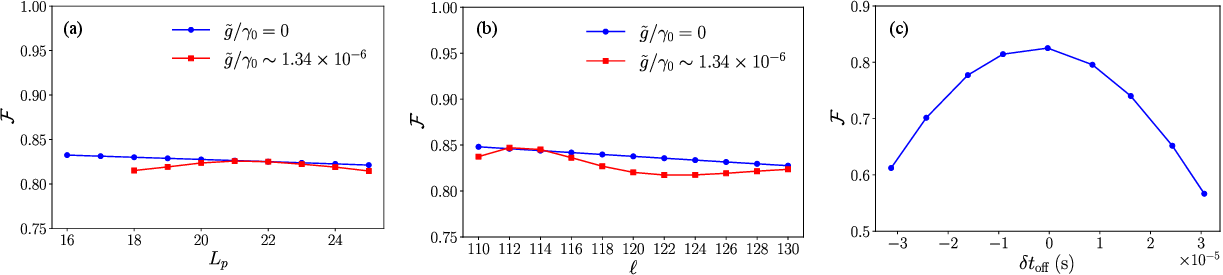}
		\caption{\label{fig:figure_a3}(a)~Fidelity versus $L_p$ without (blue) interatomic interaction, with (red) interatomic interaction (only the value of $L_{p}$ that satisfy the inequality of Eq.~(\ref{eq:constraint_1}) have been plotted. (b)~Fidelity versus $\ell$ keeping $L_p$ as constant. (c)~Fidelity versus $\delta t\textsubscript{off}$, a small variation about $t\textsubscript{off} = \pi/2\tilde{G}\sim 61$ $\mu$s. The parameters used for (a): control power$\sim8.6\times10^{-7}$ mW, $\ell = 130$, $N = 2\times10^4$ and rest same as mentioned earlier. For (b,c) everything remains with $L_p = 20$.}
	\end{figure}

	The variation of the single photon rotational eigenstate fidelity storage with respect to $L_p$, $\ell$, and $\delta t\textsubscript{off}$ while satisfying Eqs.~\eqref{eq:constraint_1} and \eqref{eq:constraint_2}, is shown in Figs.~\ref{fig:figure_a3}(a,b,c), respectively. Here, $\delta t\textsubscript{off}$ is a small variation about $t\textsubscript{off}$. In Fig.~\ref{fig:figure_a3}(a,b) it can be seen that, in the absence of interatomic interaction (blue curves), fidelity slowly decreases with $L_{p}(l)$. In the presence of interatomic interactions (red curves) there is a slight variation from the interaction free values. In Fig.~\ref{fig:figure_a3}(c), fidelity doesn't vary much if the switching off time of the control field a varies within $\pm 10$ $\mu$s around $t\textsubscript{off}$. 
	
	
	\section{Logarithmic Negativity\label{App-D}}
	The definition of logarithmic negativity $\mathcal{E}_N$ used in the main text is given by,
	\begin{equation}
		\mathcal{E}_N = \log_2||\rho^{T_a}_{a,b}||_1\label{eq:logarithmic_negativity}
	\end{equation}
	where, $\rho_{a,b}$ is the joint density matrix of the subsystems $a$ and $b$,  $\rho^{T_a}_{a,b}$ is the partial transpose of $\rho_{a,b}$ over subsystem $a$, and $||\rho^{T_a}_{a,b}||_1$ denotes the trace norm of $\rho^{T_a}_{a,b}$, which is the sum of absolute values of the eigenvalues of $\rho^{T_a}_{a,b}$. For the OAM entangled state given in Eq.~(6) of the main text, $\mathcal{E}_N = 1$ with the subsystems indexed as 1 and 2.
\twocolumngrid
	%
	\bibliographystyle{apsrev4-2}
	\bibliography{arxiv_bibliography}

\begin{thebibliography}{50}%
\makeatletter
\providecommand \@ifxundefined [1]{%
 \@ifx{#1\undefined}
}%
\providecommand \@ifnum [1]{%
 \ifnum #1\expandafter \@firstoftwo
 \else \expandafter \@secondoftwo
 \fi
}%
\providecommand \@ifx [1]{%
 \ifx #1\expandafter \@firstoftwo
 \else \expandafter \@secondoftwo
 \fi
}%
\providecommand \natexlab [1]{#1}%
\providecommand \enquote  [1]{``#1''}%
\providecommand \bibnamefont  [1]{#1}%
\providecommand \bibfnamefont [1]{#1}%
\providecommand \citenamefont [1]{#1}%
\providecommand \href@noop [0]{\@secondoftwo}%
\providecommand \href [0]{\begingroup \@sanitize@url \@href}%
\providecommand \@href[1]{\@@startlink{#1}\@@href}%
\providecommand \@@href[1]{\endgroup#1\@@endlink}%
\providecommand \@sanitize@url [0]{\catcode `\\12\catcode `\$12\catcode
  `\&12\catcode `\#12\catcode `\^12\catcode `\_12\catcode `\%12\relax}%
\providecommand \@@startlink[1]{}%
\providecommand \@@endlink[0]{}%
\providecommand \url  [0]{\begingroup\@sanitize@url \@url }%
\providecommand \@url [1]{\endgroup\@href {#1}{\urlprefix }}%
\providecommand \urlprefix  [0]{URL }%
\providecommand \Eprint [0]{\href }%
\providecommand \doibase [0]{https://doi.org/}%
\providecommand \selectlanguage [0]{\@gobble}%
\providecommand \bibinfo  [0]{\@secondoftwo}%
\providecommand \bibfield  [0]{\@secondoftwo}%
\providecommand \translation [1]{[#1]}%
\providecommand \BibitemOpen [0]{}%
\providecommand \bibitemStop [0]{}%
\providecommand \bibitemNoStop [0]{.\EOS\space}%
\providecommand \EOS [0]{\spacefactor3000\relax}%
\providecommand \BibitemShut  [1]{\csname bibitem#1\endcsname}%
\let\auto@bib@innerbib\@empty
\bibitem [{\citenamefont {Heshami}\ \emph {et~al.}(2016)\citenamefont
  {Heshami}, \citenamefont {England}, \citenamefont {Humphreys}, \citenamefont
  {Bustard}, \citenamefont {Acosta}, \citenamefont {Nunn},\ and\ \citenamefont
  {and}}]{Heshami12112016}%
  \BibitemOpen
  \bibfield  {author} {\bibinfo {author} {\bibfnamefont {K.}~\bibnamefont
  {Heshami}}, \bibinfo {author} {\bibfnamefont {D.~G.}\ \bibnamefont
  {England}}, \bibinfo {author} {\bibfnamefont {P.~C.}\ \bibnamefont
  {Humphreys}}, \bibinfo {author} {\bibfnamefont {P.~J.}\ \bibnamefont
  {Bustard}}, \bibinfo {author} {\bibfnamefont {V.~M.}\ \bibnamefont {Acosta}},
  \bibinfo {author} {\bibfnamefont {J.}~\bibnamefont {Nunn}},\ and\ \bibinfo
  {author} {\bibfnamefont {B.~J.~S.}\ \bibnamefont {and}},\ }\href
  {https://doi.org/10.1080/09500340.2016.1148212} {\bibfield  {journal}
  {\bibinfo  {journal} {J. Mod. Opt.}\ }\textbf {\bibinfo {volume} {63}},\
  \bibinfo {pages} {2005} (\bibinfo {year} {2016})}\BibitemShut {NoStop}%
\bibitem [{\citenamefont {Lvovsky}\ \emph {et~al.}(2009)\citenamefont
  {Lvovsky}, \citenamefont {Sanders},\ and\ \citenamefont
  {Tittel}}]{lvovsky_optical_2009}%
  \BibitemOpen
  \bibfield  {author} {\bibinfo {author} {\bibfnamefont {A.~I.}\ \bibnamefont
  {Lvovsky}}, \bibinfo {author} {\bibfnamefont {B.~C.}\ \bibnamefont
  {Sanders}},\ and\ \bibinfo {author} {\bibfnamefont {W.}~\bibnamefont
  {Tittel}},\ }\href {https://doi.org/10.1038/nphoton.2009.231} {\bibfield
  {journal} {\bibinfo  {journal} {Nat. Photonics}\ }\textbf {\bibinfo {volume}
  {3}},\ \bibinfo {pages} {706} (\bibinfo {year} {2009})}\BibitemShut {NoStop}%
\bibitem [{\citenamefont {Veissier}(2013)}]{veissier:tel-00977307}%
  \BibitemOpen
  \bibfield  {author} {\bibinfo {author} {\bibfnamefont {L.}~\bibnamefont
  {Veissier}},\ }\emph {\bibinfo {title} {{Quantum memory protocols in large
  cold atomic ensembles}}},\ \href {https://theses.hal.science/tel-00977307}
  {\bibinfo {type} {Ph. d. thesis}},\ \bibinfo  {school} {{Universit{\'e}
  Pierre et Marie Curie - Paris VI}} (\bibinfo {year} {2013})\BibitemShut
  {NoStop}%
\bibitem [{\citenamefont {He}\ \emph {et~al.}(2022)\citenamefont {He},
  \citenamefont {Shen},\ and\ \citenamefont {Forbes}}]{he_towards_2022}%
  \BibitemOpen
  \bibfield  {author} {\bibinfo {author} {\bibfnamefont {C.}~\bibnamefont
  {He}}, \bibinfo {author} {\bibfnamefont {Y.}~\bibnamefont {Shen}},\ and\
  \bibinfo {author} {\bibfnamefont {A.}~\bibnamefont {Forbes}},\ }\href
  {https://doi.org/10.1038/s41377-022-00897-3} {\bibfield  {journal} {\bibinfo
  {journal} {Light: Science \& Applications}\ }\textbf {\bibinfo {volume}
  {11}},\ \bibinfo {pages} {205} (\bibinfo {year} {2022})}\BibitemShut
  {NoStop}%
\bibitem [{\citenamefont {Shi}\ \emph {et~al.}(2018)\citenamefont {Shi},
  \citenamefont {Ding},\ and\ \citenamefont {Zhang}}]{Shi_2018}%
  \BibitemOpen
  \bibfield  {author} {\bibinfo {author} {\bibfnamefont {B.-S.}\ \bibnamefont
  {Shi}}, \bibinfo {author} {\bibfnamefont {D.-S.}\ \bibnamefont {Ding}},\ and\
  \bibinfo {author} {\bibfnamefont {W.}~\bibnamefont {Zhang}},\ }\href
  {https://doi.org/10.1088/1361-6455/aa9b95} {\bibfield  {journal} {\bibinfo
  {journal} {J. Phys. B: Atomic, Molecular and Optical Physics}\ }\textbf
  {\bibinfo {volume} {51}},\ \bibinfo {pages} {032004} (\bibinfo {year}
  {2018})}\BibitemShut {NoStop}%
\bibitem [{\citenamefont {Ding}\ \emph {et~al.}(2016)\citenamefont {Ding},
  \citenamefont {Zhang}, \citenamefont {Shi}, \citenamefont {Zhou},
  \citenamefont {Li}, \citenamefont {Shi},\ and\ \citenamefont
  {Guo}}]{ding_high-dimensional_2016}%
  \BibitemOpen
  \bibfield  {author} {\bibinfo {author} {\bibfnamefont {D.-S.}\ \bibnamefont
  {Ding}}, \bibinfo {author} {\bibfnamefont {W.}~\bibnamefont {Zhang}},
  \bibinfo {author} {\bibfnamefont {S.}~\bibnamefont {Shi}}, \bibinfo {author}
  {\bibfnamefont {Z.-Y.}\ \bibnamefont {Zhou}}, \bibinfo {author}
  {\bibfnamefont {Y.}~\bibnamefont {Li}}, \bibinfo {author} {\bibfnamefont
  {B.-S.}\ \bibnamefont {Shi}},\ and\ \bibinfo {author} {\bibfnamefont {G.-C.}\
  \bibnamefont {Guo}},\ }\href {https://doi.org/10.1038/lsa.2016.157}
  {\bibfield  {journal} {\bibinfo  {journal} {Light Sci. Appl.}\ }\textbf
  {\bibinfo {volume} {5}},\ \bibinfo {pages} {e16157} (\bibinfo {year}
  {2016})}\BibitemShut {NoStop}%
\bibitem [{\citenamefont {Ding}\ \emph {et~al.}(2015)\citenamefont {Ding},
  \citenamefont {Zhang}, \citenamefont {Zhou}, \citenamefont {Shi},
  \citenamefont {Xiang}, \citenamefont {Wang}, \citenamefont {Jiang},
  \citenamefont {Shi},\ and\ \citenamefont {Guo}}]{PhysRevLett.114.050502}%
  \BibitemOpen
  \bibfield  {author} {\bibinfo {author} {\bibfnamefont {D.-S.}\ \bibnamefont
  {Ding}}, \bibinfo {author} {\bibfnamefont {W.}~\bibnamefont {Zhang}},
  \bibinfo {author} {\bibfnamefont {Z.-Y.}\ \bibnamefont {Zhou}}, \bibinfo
  {author} {\bibfnamefont {S.}~\bibnamefont {Shi}}, \bibinfo {author}
  {\bibfnamefont {G.-Y.}\ \bibnamefont {Xiang}}, \bibinfo {author}
  {\bibfnamefont {X.-S.}\ \bibnamefont {Wang}}, \bibinfo {author}
  {\bibfnamefont {Y.-K.}\ \bibnamefont {Jiang}}, \bibinfo {author}
  {\bibfnamefont {B.-S.}\ \bibnamefont {Shi}},\ and\ \bibinfo {author}
  {\bibfnamefont {G.-C.}\ \bibnamefont {Guo}},\ }\href
  {https://doi.org/10.1103/PhysRevLett.114.050502} {\bibfield  {journal}
  {\bibinfo  {journal} {Phys. Rev. Lett.}\ }\textbf {\bibinfo {volume} {114}},\
  \bibinfo {pages} {050502} (\bibinfo {year} {2015})}\BibitemShut {NoStop}%
\bibitem [{\citenamefont {Nicolas}\ \emph {et~al.}(2014)\citenamefont
  {Nicolas}, \citenamefont {Veissier}, \citenamefont {Giner}, \citenamefont
  {Giacobino}, \citenamefont {Maxein},\ and\ \citenamefont
  {Laurat}}]{nicolas_quantum_2014}%
  \BibitemOpen
  \bibfield  {author} {\bibinfo {author} {\bibfnamefont {A.}~\bibnamefont
  {Nicolas}}, \bibinfo {author} {\bibfnamefont {L.}~\bibnamefont {Veissier}},
  \bibinfo {author} {\bibfnamefont {L.}~\bibnamefont {Giner}}, \bibinfo
  {author} {\bibfnamefont {E.}~\bibnamefont {Giacobino}}, \bibinfo {author}
  {\bibfnamefont {D.}~\bibnamefont {Maxein}},\ and\ \bibinfo {author}
  {\bibfnamefont {J.}~\bibnamefont {Laurat}},\ }\href
  {https://doi.org/10.1038/nphoton.2013.355} {\bibfield  {journal} {\bibinfo
  {journal} {Nat. Photonics}\ }\textbf {\bibinfo {volume} {8}},\ \bibinfo
  {pages} {234} (\bibinfo {year} {2014})}\BibitemShut {NoStop}%
\bibitem [{\citenamefont {Moretti}\ \emph {et~al.}(2009)\citenamefont
  {Moretti}, \citenamefont {Felinto},\ and\ \citenamefont
  {Tabosa}}]{PhysRevA.79.023825}%
  \BibitemOpen
  \bibfield  {author} {\bibinfo {author} {\bibfnamefont {D.}~\bibnamefont
  {Moretti}}, \bibinfo {author} {\bibfnamefont {D.}~\bibnamefont {Felinto}},\
  and\ \bibinfo {author} {\bibfnamefont {J.~W.~R.}\ \bibnamefont {Tabosa}},\
  }\href {https://doi.org/10.1103/PhysRevA.79.023825} {\bibfield  {journal}
  {\bibinfo  {journal} {Phys. Rev. A}\ }\textbf {\bibinfo {volume} {79}},\
  \bibinfo {pages} {023825} (\bibinfo {year} {2009})}\BibitemShut {NoStop}%
\bibitem [{\citenamefont {Veissier}\ \emph {et~al.}(2013)\citenamefont
  {Veissier}, \citenamefont {Nicolas}, \citenamefont {Giner}, \citenamefont
  {Maxein}, \citenamefont {Sheremet}, \citenamefont {Giacobino},\ and\
  \citenamefont {Laurat}}]{Veissier:13}%
  \BibitemOpen
  \bibfield  {author} {\bibinfo {author} {\bibfnamefont {L.}~\bibnamefont
  {Veissier}}, \bibinfo {author} {\bibfnamefont {A.}~\bibnamefont {Nicolas}},
  \bibinfo {author} {\bibfnamefont {L.}~\bibnamefont {Giner}}, \bibinfo
  {author} {\bibfnamefont {D.}~\bibnamefont {Maxein}}, \bibinfo {author}
  {\bibfnamefont {A.~S.}\ \bibnamefont {Sheremet}}, \bibinfo {author}
  {\bibfnamefont {E.}~\bibnamefont {Giacobino}},\ and\ \bibinfo {author}
  {\bibfnamefont {J.}~\bibnamefont {Laurat}},\ }\href
  {https://doi.org/10.1364/OL.38.000712} {\bibfield  {journal} {\bibinfo
  {journal} {Opt. Lett.}\ }\textbf {\bibinfo {volume} {38}},\ \bibinfo {pages}
  {712} (\bibinfo {year} {2013})}\BibitemShut {NoStop}%
\bibitem [{\citenamefont {Wang}\ \emph {et~al.}(2021)\citenamefont {Wang},
  \citenamefont {Yu}, \citenamefont {Chen}, \citenamefont {Cao}, \citenamefont
  {Wang}, \citenamefont {Yang}, \citenamefont {Qiu}, \citenamefont {Wei},
  \citenamefont {Gao},\ and\ \citenamefont {Li}}]{Wang_2021}%
  \BibitemOpen
  \bibfield  {author} {\bibinfo {author} {\bibfnamefont {C.}~\bibnamefont
  {Wang}}, \bibinfo {author} {\bibfnamefont {Y.}~\bibnamefont {Yu}}, \bibinfo
  {author} {\bibfnamefont {Y.}~\bibnamefont {Chen}}, \bibinfo {author}
  {\bibfnamefont {M.}~\bibnamefont {Cao}}, \bibinfo {author} {\bibfnamefont
  {J.}~\bibnamefont {Wang}}, \bibinfo {author} {\bibfnamefont {X.}~\bibnamefont
  {Yang}}, \bibinfo {author} {\bibfnamefont {S.}~\bibnamefont {Qiu}}, \bibinfo
  {author} {\bibfnamefont {D.}~\bibnamefont {Wei}}, \bibinfo {author}
  {\bibfnamefont {H.}~\bibnamefont {Gao}},\ and\ \bibinfo {author}
  {\bibfnamefont {F.}~\bibnamefont {Li}},\ }\href
  {https://doi.org/10.1088/2058-9565/ac120a} {\bibfield  {journal} {\bibinfo
  {journal} {Quantum Science and Technology}\ }\textbf {\bibinfo {volume}
  {6}},\ \bibinfo {pages} {045008} (\bibinfo {year} {2021})}\BibitemShut
  {NoStop}%
\bibitem [{\citenamefont {Zhou}\ \emph {et~al.}(2015)\citenamefont {Zhou},
  \citenamefont {Hua}, \citenamefont {Liu}, \citenamefont {Chen}, \citenamefont
  {Xu}, \citenamefont {Han}, \citenamefont {Li},\ and\ \citenamefont
  {Guo}}]{PhysRevLett.115.070502}%
  \BibitemOpen
  \bibfield  {author} {\bibinfo {author} {\bibfnamefont {Z.-Q.}\ \bibnamefont
  {Zhou}}, \bibinfo {author} {\bibfnamefont {Y.-L.}\ \bibnamefont {Hua}},
  \bibinfo {author} {\bibfnamefont {X.}~\bibnamefont {Liu}}, \bibinfo {author}
  {\bibfnamefont {G.}~\bibnamefont {Chen}}, \bibinfo {author} {\bibfnamefont
  {J.-S.}\ \bibnamefont {Xu}}, \bibinfo {author} {\bibfnamefont {Y.-J.}\
  \bibnamefont {Han}}, \bibinfo {author} {\bibfnamefont {C.-F.}\ \bibnamefont
  {Li}},\ and\ \bibinfo {author} {\bibfnamefont {G.-C.}\ \bibnamefont {Guo}},\
  }\href {https://doi.org/10.1103/PhysRevLett.115.070502} {\bibfield  {journal}
  {\bibinfo  {journal} {Phys. Rev. Lett.}\ }\textbf {\bibinfo {volume} {115}},\
  \bibinfo {pages} {070502} (\bibinfo {year} {2015})}\BibitemShut {NoStop}%
\bibitem [{\citenamefont {Ye}\ \emph {et~al.}(2022)\citenamefont {Ye},
  \citenamefont {Zeng}, \citenamefont {Dong}, \citenamefont {Zhang},
  \citenamefont {Li}, \citenamefont {Li}, \citenamefont {Guo}, \citenamefont
  {Ding},\ and\ \citenamefont {Shi}}]{PhysRevLett.129.193601}%
  \BibitemOpen
  \bibfield  {author} {\bibinfo {author} {\bibfnamefont {Y.-H.}\ \bibnamefont
  {Ye}}, \bibinfo {author} {\bibfnamefont {L.}~\bibnamefont {Zeng}}, \bibinfo
  {author} {\bibfnamefont {M.-X.}\ \bibnamefont {Dong}}, \bibinfo {author}
  {\bibfnamefont {W.-H.}\ \bibnamefont {Zhang}}, \bibinfo {author}
  {\bibfnamefont {E.-Z.}\ \bibnamefont {Li}}, \bibinfo {author} {\bibfnamefont
  {D.-C.}\ \bibnamefont {Li}}, \bibinfo {author} {\bibfnamefont {G.-C.}\
  \bibnamefont {Guo}}, \bibinfo {author} {\bibfnamefont {D.-S.}\ \bibnamefont
  {Ding}},\ and\ \bibinfo {author} {\bibfnamefont {B.-S.}\ \bibnamefont
  {Shi}},\ }\href {https://doi.org/10.1103/PhysRevLett.129.193601} {\bibfield
  {journal} {\bibinfo  {journal} {Phys. Rev. Lett.}\ }\textbf {\bibinfo
  {volume} {129}},\ \bibinfo {pages} {193601} (\bibinfo {year}
  {2022})}\BibitemShut {NoStop}%
\bibitem [{\citenamefont {Leggett}(2001)}]{leggett_bose-einstein_2001}%
  \BibitemOpen
  \bibfield  {author} {\bibinfo {author} {\bibfnamefont {A.~J.}\ \bibnamefont
  {Leggett}},\ }\href {https://doi.org/10.1103/RevModPhys.73.307} {\bibfield
  {journal} {\bibinfo  {journal} {Rev. Mod. Phys}\ }\textbf {\bibinfo {volume}
  {73}},\ \bibinfo {pages} {307} (\bibinfo {year} {2001})}\BibitemShut
  {NoStop}%
\bibitem [{\citenamefont {Morizot}\ \emph {et~al.}(2006)\citenamefont
  {Morizot}, \citenamefont {Colombe}, \citenamefont {Lorent}, \citenamefont
  {Perrin},\ and\ \citenamefont {Garraway}}]{PhysRevA.74.023617}%
  \BibitemOpen
  \bibfield  {author} {\bibinfo {author} {\bibfnamefont {O.}~\bibnamefont
  {Morizot}}, \bibinfo {author} {\bibfnamefont {Y.}~\bibnamefont {Colombe}},
  \bibinfo {author} {\bibfnamefont {V.}~\bibnamefont {Lorent}}, \bibinfo
  {author} {\bibfnamefont {H.}~\bibnamefont {Perrin}},\ and\ \bibinfo {author}
  {\bibfnamefont {B.~M.}\ \bibnamefont {Garraway}},\ }\href
  {https://doi.org/10.1103/PhysRevA.74.023617} {\bibfield  {journal} {\bibinfo
  {journal} {Phys. Rev. A}\ }\textbf {\bibinfo {volume} {74}},\ \bibinfo
  {pages} {023617} (\bibinfo {year} {2006})}\BibitemShut {NoStop}%
\bibitem [{\citenamefont {Sherlock}\ \emph {et~al.}(2011)\citenamefont
  {Sherlock}, \citenamefont {Gildemeister}, \citenamefont {Owen}, \citenamefont
  {Nugent},\ and\ \citenamefont {Foot}}]{PhysRevA.83.043408}%
  \BibitemOpen
  \bibfield  {author} {\bibinfo {author} {\bibfnamefont {B.~E.}\ \bibnamefont
  {Sherlock}}, \bibinfo {author} {\bibfnamefont {M.}~\bibnamefont
  {Gildemeister}}, \bibinfo {author} {\bibfnamefont {E.}~\bibnamefont {Owen}},
  \bibinfo {author} {\bibfnamefont {E.}~\bibnamefont {Nugent}},\ and\ \bibinfo
  {author} {\bibfnamefont {C.~J.}\ \bibnamefont {Foot}},\ }\href
  {https://doi.org/10.1103/PhysRevA.83.043408} {\bibfield  {journal} {\bibinfo
  {journal} {Phys. Rev. A}\ }\textbf {\bibinfo {volume} {83}},\ \bibinfo
  {pages} {043408} (\bibinfo {year} {2011})}\BibitemShut {NoStop}%
\bibitem [{\citenamefont {Ryu}\ \emph {et~al.}(2007)\citenamefont {Ryu},
  \citenamefont {Andersen}, \citenamefont {Clad\'e}, \citenamefont {Natarajan},
  \citenamefont {Helmerson},\ and\ \citenamefont
  {Phillips}}]{PhysRevLett.99.260401}%
  \BibitemOpen
  \bibfield  {author} {\bibinfo {author} {\bibfnamefont {C.}~\bibnamefont
  {Ryu}}, \bibinfo {author} {\bibfnamefont {M.~F.}\ \bibnamefont {Andersen}},
  \bibinfo {author} {\bibfnamefont {P.}~\bibnamefont {Clad\'e}}, \bibinfo
  {author} {\bibfnamefont {V.}~\bibnamefont {Natarajan}}, \bibinfo {author}
  {\bibfnamefont {K.}~\bibnamefont {Helmerson}},\ and\ \bibinfo {author}
  {\bibfnamefont {W.~D.}\ \bibnamefont {Phillips}},\ }\href
  {https://doi.org/10.1103/PhysRevLett.99.260401} {\bibfield  {journal}
  {\bibinfo  {journal} {Phys. Rev. Lett.}\ }\textbf {\bibinfo {volume} {99}},\
  \bibinfo {pages} {260401} (\bibinfo {year} {2007})}\BibitemShut {NoStop}%
\bibitem [{\citenamefont {Wright}\ \emph {et~al.}(2013)\citenamefont {Wright},
  \citenamefont {Blakestad}, \citenamefont {Lobb}, \citenamefont {Phillips},\
  and\ \citenamefont {Campbell}}]{PhysRevLett.110.025302}%
  \BibitemOpen
  \bibfield  {author} {\bibinfo {author} {\bibfnamefont {K.~C.}\ \bibnamefont
  {Wright}}, \bibinfo {author} {\bibfnamefont {R.~B.}\ \bibnamefont
  {Blakestad}}, \bibinfo {author} {\bibfnamefont {C.~J.}\ \bibnamefont {Lobb}},
  \bibinfo {author} {\bibfnamefont {W.~D.}\ \bibnamefont {Phillips}},\ and\
  \bibinfo {author} {\bibfnamefont {G.~K.}\ \bibnamefont {Campbell}},\ }\href
  {https://doi.org/10.1103/PhysRevLett.110.025302} {\bibfield  {journal}
  {\bibinfo  {journal} {Phys. Rev. Lett.}\ }\textbf {\bibinfo {volume} {110}},\
  \bibinfo {pages} {025302} (\bibinfo {year} {2013})}\BibitemShut {NoStop}%
\bibitem [{\citenamefont {Guo}\ \emph {et~al.}(2020)\citenamefont {Guo},
  \citenamefont {Dubessy}, \citenamefont {de~Herve}, \citenamefont {Kumar},
  \citenamefont {Badr}, \citenamefont {Perrin}, \citenamefont {Longchambon},\
  and\ \citenamefont {Perrin}}]{PhysRevLett.124.025301}%
  \BibitemOpen
  \bibfield  {author} {\bibinfo {author} {\bibfnamefont {Y.}~\bibnamefont
  {Guo}}, \bibinfo {author} {\bibfnamefont {R.}~\bibnamefont {Dubessy}},
  \bibinfo {author} {\bibfnamefont {M.~d.~G.}\ \bibnamefont {de~Herve}},
  \bibinfo {author} {\bibfnamefont {A.}~\bibnamefont {Kumar}}, \bibinfo
  {author} {\bibfnamefont {T.}~\bibnamefont {Badr}}, \bibinfo {author}
  {\bibfnamefont {A.}~\bibnamefont {Perrin}}, \bibinfo {author} {\bibfnamefont
  {L.}~\bibnamefont {Longchambon}},\ and\ \bibinfo {author} {\bibfnamefont
  {H.}~\bibnamefont {Perrin}},\ }\href
  {https://doi.org/10.1103/PhysRevLett.124.025301} {\bibfield  {journal}
  {\bibinfo  {journal} {Phys. Rev. Lett.}\ }\textbf {\bibinfo {volume} {124}},\
  \bibinfo {pages} {025301} (\bibinfo {year} {2020})}\BibitemShut {NoStop}%
\bibitem [{\citenamefont {Ramanathan}\ \emph {et~al.}(2011)\citenamefont
  {Ramanathan}, \citenamefont {Wright}, \citenamefont {Muniz}, \citenamefont
  {Zelan}, \citenamefont {Hill}, \citenamefont {Lobb}, \citenamefont
  {Helmerson}, \citenamefont {Phillips},\ and\ \citenamefont
  {Campbell}}]{PhysRevLett.106.130401}%
  \BibitemOpen
  \bibfield  {author} {\bibinfo {author} {\bibfnamefont {A.}~\bibnamefont
  {Ramanathan}}, \bibinfo {author} {\bibfnamefont {K.~C.}\ \bibnamefont
  {Wright}}, \bibinfo {author} {\bibfnamefont {S.~R.}\ \bibnamefont {Muniz}},
  \bibinfo {author} {\bibfnamefont {M.}~\bibnamefont {Zelan}}, \bibinfo
  {author} {\bibfnamefont {W.~T.}\ \bibnamefont {Hill}}, \bibinfo {author}
  {\bibfnamefont {C.~J.}\ \bibnamefont {Lobb}}, \bibinfo {author}
  {\bibfnamefont {K.}~\bibnamefont {Helmerson}}, \bibinfo {author}
  {\bibfnamefont {W.~D.}\ \bibnamefont {Phillips}},\ and\ \bibinfo {author}
  {\bibfnamefont {G.~K.}\ \bibnamefont {Campbell}},\ }\href
  {https://doi.org/10.1103/PhysRevLett.106.130401} {\bibfield  {journal}
  {\bibinfo  {journal} {Phys. Rev. Lett.}\ }\textbf {\bibinfo {volume} {106}},\
  \bibinfo {pages} {130401} (\bibinfo {year} {2011})}\BibitemShut {NoStop}%
\bibitem [{\citenamefont {Beattie}\ \emph {et~al.}(2013)\citenamefont
  {Beattie}, \citenamefont {Moulder}, \citenamefont {Fletcher},\ and\
  \citenamefont {Hadzibabic}}]{PhysRevLett.110.025301}%
  \BibitemOpen
  \bibfield  {author} {\bibinfo {author} {\bibfnamefont {S.}~\bibnamefont
  {Beattie}}, \bibinfo {author} {\bibfnamefont {S.}~\bibnamefont {Moulder}},
  \bibinfo {author} {\bibfnamefont {R.~J.}\ \bibnamefont {Fletcher}},\ and\
  \bibinfo {author} {\bibfnamefont {Z.}~\bibnamefont {Hadzibabic}},\ }\href
  {https://doi.org/10.1103/PhysRevLett.110.025301} {\bibfield  {journal}
  {\bibinfo  {journal} {Phys. Rev. Lett.}\ }\textbf {\bibinfo {volume} {110}},\
  \bibinfo {pages} {025301} (\bibinfo {year} {2013})}\BibitemShut {NoStop}%
\bibitem [{\citenamefont {Kapale}\ and\ \citenamefont
  {Dowling}(2005)}]{kapale_dowling_vortex_phase_qubitPRL2005}%
  \BibitemOpen
  \bibfield  {author} {\bibinfo {author} {\bibfnamefont {K.~T.}\ \bibnamefont
  {Kapale}}\ and\ \bibinfo {author} {\bibfnamefont {J.~P.}\ \bibnamefont
  {Dowling}},\ }\href {https://doi.org/10.1103/PhysRevLett.95.173601}
  {\bibfield  {journal} {\bibinfo  {journal} {Phys. Rev. Lett.}\ }\textbf
  {\bibinfo {volume} {95}},\ \bibinfo {pages} {173601} (\bibinfo {year}
  {2005})}\BibitemShut {NoStop}%
\bibitem [{\citenamefont {Pandey}\ \emph {et~al.}(2019)\citenamefont {Pandey},
  \citenamefont {Mas}, \citenamefont {Drougakis}, \citenamefont {Thekkeppatt},
  \citenamefont {Bolpasi}, \citenamefont {Vasilakis}, \citenamefont {Poulios},\
  and\ \citenamefont {von Klitzing}}]{von_Klitzing_hypersonic_BEC}%
  \BibitemOpen
  \bibfield  {author} {\bibinfo {author} {\bibfnamefont {S.}~\bibnamefont
  {Pandey}}, \bibinfo {author} {\bibfnamefont {H.}~\bibnamefont {Mas}},
  \bibinfo {author} {\bibfnamefont {G.}~\bibnamefont {Drougakis}}, \bibinfo
  {author} {\bibfnamefont {P.}~\bibnamefont {Thekkeppatt}}, \bibinfo {author}
  {\bibfnamefont {V.}~\bibnamefont {Bolpasi}}, \bibinfo {author} {\bibfnamefont
  {G.}~\bibnamefont {Vasilakis}}, \bibinfo {author} {\bibfnamefont
  {K.}~\bibnamefont {Poulios}},\ and\ \bibinfo {author} {\bibfnamefont {W.~v.}\
  \bibnamefont {von Klitzing}},\ }\href
  {https://doi.org/https://doi.org/10.1038/s41586-019-1273-5} {\bibfield
  {journal} {\bibinfo  {journal} {Nature}\ }\textbf {\bibinfo {volume} {570}},\
  \bibinfo {pages} {205} (\bibinfo {year} {2019})}\BibitemShut {NoStop}%
\bibitem [{\citenamefont {Tian}\ and\ \citenamefont
  {Wang}(2010)}]{PhysRevA.82.053806}%
  \BibitemOpen
  \bibfield  {author} {\bibinfo {author} {\bibfnamefont {L.}~\bibnamefont
  {Tian}}\ and\ \bibinfo {author} {\bibfnamefont {H.}~\bibnamefont {Wang}},\
  }\href {https://doi.org/10.1103/PhysRevA.82.053806} {\bibfield  {journal}
  {\bibinfo  {journal} {Phys. Rev. A}\ }\textbf {\bibinfo {volume} {82}},\
  \bibinfo {pages} {053806} (\bibinfo {year} {2010})}\BibitemShut {NoStop}%
\bibitem [{\citenamefont {Kaviani}\ \emph {et~al.}(2020)\citenamefont
  {Kaviani}, \citenamefont {Ghobadi}, \citenamefont {Behera}, \citenamefont
  {Wu}, \citenamefont {Hryciw}, \citenamefont {Vo}, \citenamefont {Fattal},\
  and\ \citenamefont {Barclay}}]{Kaviani:20}%
  \BibitemOpen
  \bibfield  {author} {\bibinfo {author} {\bibfnamefont {H.}~\bibnamefont
  {Kaviani}}, \bibinfo {author} {\bibfnamefont {R.}~\bibnamefont {Ghobadi}},
  \bibinfo {author} {\bibfnamefont {B.}~\bibnamefont {Behera}}, \bibinfo
  {author} {\bibfnamefont {M.}~\bibnamefont {Wu}}, \bibinfo {author}
  {\bibfnamefont {A.}~\bibnamefont {Hryciw}}, \bibinfo {author} {\bibfnamefont
  {S.}~\bibnamefont {Vo}}, \bibinfo {author} {\bibfnamefont {D.}~\bibnamefont
  {Fattal}},\ and\ \bibinfo {author} {\bibfnamefont {P.}~\bibnamefont
  {Barclay}},\ }\href {https://doi.org/10.1364/OE.389170} {\bibfield  {journal}
  {\bibinfo  {journal} {Opt. Express}\ }\textbf {\bibinfo {volume} {28}},\
  \bibinfo {pages} {15482} (\bibinfo {year} {2020})}\BibitemShut {NoStop}%
\bibitem [{\citenamefont {Leach}\ \emph {et~al.}(2002)\citenamefont {Leach},
  \citenamefont {Padgett}, \citenamefont {Barnett}, \citenamefont
  {Franke-Arnold},\ and\ \citenamefont {Courtial}}]{PhysRevLett.88.257901}%
  \BibitemOpen
  \bibfield  {author} {\bibinfo {author} {\bibfnamefont {J.}~\bibnamefont
  {Leach}}, \bibinfo {author} {\bibfnamefont {M.~J.}\ \bibnamefont {Padgett}},
  \bibinfo {author} {\bibfnamefont {S.~M.}\ \bibnamefont {Barnett}}, \bibinfo
  {author} {\bibfnamefont {S.}~\bibnamefont {Franke-Arnold}},\ and\ \bibinfo
  {author} {\bibfnamefont {J.}~\bibnamefont {Courtial}},\ }\href
  {https://doi.org/10.1103/PhysRevLett.88.257901} {\bibfield  {journal}
  {\bibinfo  {journal} {Phys. Rev. Lett.}\ }\textbf {\bibinfo {volume} {88}},\
  \bibinfo {pages} {257901} (\bibinfo {year} {2002})}\BibitemShut {NoStop}%
\bibitem [{\citenamefont {Yao}\ and\ \citenamefont {Padgett}(2011)}]{Yao:11}%
  \BibitemOpen
  \bibfield  {author} {\bibinfo {author} {\bibfnamefont {A.~M.}\ \bibnamefont
  {Yao}}\ and\ \bibinfo {author} {\bibfnamefont {M.~J.}\ \bibnamefont
  {Padgett}},\ }\href {https://doi.org/10.1364/AOP.3.000161} {\bibfield
  {journal} {\bibinfo  {journal} {Adv. Opt. Photon.}\ }\textbf {\bibinfo
  {volume} {3}},\ \bibinfo {pages} {161} (\bibinfo {year} {2011})}\BibitemShut
  {NoStop}%
\bibitem [{\citenamefont {Kumar}\ \emph {et~al.}(2021)\citenamefont {Kumar},
  \citenamefont {Biswas}, \citenamefont {Feliz}, \citenamefont {Kanamoto},
  \citenamefont {Chang}, \citenamefont {Jha},\ and\ \citenamefont
  {Bhattacharya}}]{kumar_cavity_2021}%
  \BibitemOpen
  \bibfield  {author} {\bibinfo {author} {\bibfnamefont {P.}~\bibnamefont
  {Kumar}}, \bibinfo {author} {\bibfnamefont {T.}~\bibnamefont {Biswas}},
  \bibinfo {author} {\bibfnamefont {K.}~\bibnamefont {Feliz}}, \bibinfo
  {author} {\bibfnamefont {R.}~\bibnamefont {Kanamoto}}, \bibinfo {author}
  {\bibfnamefont {M.-S.}\ \bibnamefont {Chang}}, \bibinfo {author}
  {\bibfnamefont {A.~K.}\ \bibnamefont {Jha}},\ and\ \bibinfo {author}
  {\bibfnamefont {M.}~\bibnamefont {Bhattacharya}},\ }\href
  {https://doi.org/10.1103/PhysRevLett.127.113601} {\bibfield  {journal}
  {\bibinfo  {journal} {Phys. Rev. Lett.}\ }\textbf {\bibinfo {volume} {127}},\
  \bibinfo {pages} {113601} (\bibinfo {year} {2021})}\BibitemShut {NoStop}%
\bibitem [{\citenamefont {Aspelmeyer}\ \emph {et~al.}(2014)\citenamefont
  {Aspelmeyer}, \citenamefont {Kippenberg},\ and\ \citenamefont
  {Marquardt}}]{aspelmeyer_cavity_2014}%
  \BibitemOpen
  \bibfield  {author} {\bibinfo {author} {\bibfnamefont {M.}~\bibnamefont
  {Aspelmeyer}}, \bibinfo {author} {\bibfnamefont {T.~J.}\ \bibnamefont
  {Kippenberg}},\ and\ \bibinfo {author} {\bibfnamefont {F.}~\bibnamefont
  {Marquardt}},\ }\href {https://doi.org/10.1103/RevModPhys.86.1391} {\bibfield
   {journal} {\bibinfo  {journal} {Rev. Mod. Phys.}\ }\textbf {\bibinfo
  {volume} {86}},\ \bibinfo {pages} {1391} (\bibinfo {year}
  {2014})}\BibitemShut {NoStop}%
\bibitem [{\citenamefont {Fiore}\ \emph {et~al.}(2011)\citenamefont {Fiore},
  \citenamefont {Yang}, \citenamefont {Kuzyk}, \citenamefont {Barbour},
  \citenamefont {Tian},\ and\ \citenamefont {Wang}}]{fiore_storing_2011}%
  \BibitemOpen
  \bibfield  {author} {\bibinfo {author} {\bibfnamefont {V.}~\bibnamefont
  {Fiore}}, \bibinfo {author} {\bibfnamefont {Y.}~\bibnamefont {Yang}},
  \bibinfo {author} {\bibfnamefont {M.~C.}\ \bibnamefont {Kuzyk}}, \bibinfo
  {author} {\bibfnamefont {R.}~\bibnamefont {Barbour}}, \bibinfo {author}
  {\bibfnamefont {L.}~\bibnamefont {Tian}},\ and\ \bibinfo {author}
  {\bibfnamefont {H.}~\bibnamefont {Wang}},\ }\href
  {https://doi.org/10.1103/PhysRevLett.107.133601} {\bibfield  {journal}
  {\bibinfo  {journal} {Phys. Rev. Lett.}\ }\textbf {\bibinfo {volume} {107}},\
  \bibinfo {pages} {133601} (\bibinfo {year} {2011})}\BibitemShut {NoStop}%
\bibitem [{\citenamefont {Weis}\ \emph {et~al.}(2010)\citenamefont {Weis},
  \citenamefont {Rivière}, \citenamefont {Deléglise}, \citenamefont
  {Gavartin}, \citenamefont {Arcizet}, \citenamefont {Schliesser},\ and\
  \citenamefont {Kippenberg}}]{doi:10.1126/science.1195596}%
  \BibitemOpen
  \bibfield  {author} {\bibinfo {author} {\bibfnamefont {S.}~\bibnamefont
  {Weis}}, \bibinfo {author} {\bibfnamefont {R.}~\bibnamefont {Rivière}},
  \bibinfo {author} {\bibfnamefont {S.}~\bibnamefont {Deléglise}}, \bibinfo
  {author} {\bibfnamefont {E.}~\bibnamefont {Gavartin}}, \bibinfo {author}
  {\bibfnamefont {O.}~\bibnamefont {Arcizet}}, \bibinfo {author} {\bibfnamefont
  {A.}~\bibnamefont {Schliesser}},\ and\ \bibinfo {author} {\bibfnamefont
  {T.~J.}\ \bibnamefont {Kippenberg}},\ }\href
  {https://doi.org/10.1126/science.1195596} {\bibfield  {journal} {\bibinfo
  {journal} {Science}\ }\textbf {\bibinfo {volume} {330}},\ \bibinfo {pages}
  {1520} (\bibinfo {year} {2010})}\BibitemShut {NoStop}%
\bibitem [{\citenamefont {Kalita}\ \emph {et~al.}(2023)\citenamefont {Kalita},
  \citenamefont {Kumar}, \citenamefont {Kanamoto}, \citenamefont
  {Bhattacharya},\ and\ \citenamefont {Sarma}}]{PhysRevA.107.013525}%
  \BibitemOpen
  \bibfield  {author} {\bibinfo {author} {\bibfnamefont {S.}~\bibnamefont
  {Kalita}}, \bibinfo {author} {\bibfnamefont {P.}~\bibnamefont {Kumar}},
  \bibinfo {author} {\bibfnamefont {R.}~\bibnamefont {Kanamoto}}, \bibinfo
  {author} {\bibfnamefont {M.}~\bibnamefont {Bhattacharya}},\ and\ \bibinfo
  {author} {\bibfnamefont {A.~K.}\ \bibnamefont {Sarma}},\ }\href
  {https://doi.org/10.1103/PhysRevA.107.013525} {\bibfield  {journal} {\bibinfo
   {journal} {Phys. Rev. A}\ }\textbf {\bibinfo {volume} {107}},\ \bibinfo
  {pages} {013525} (\bibinfo {year} {2023})}\BibitemShut {NoStop}%
\bibitem [{\citenamefont {Gardiner}\ and\ \citenamefont
  {Zoller}(2004)}]{Quantum_Noise}%
  \BibitemOpen
  \bibfield  {author} {\bibinfo {author} {\bibfnamefont {C.}~\bibnamefont
  {Gardiner}}\ and\ \bibinfo {author} {\bibfnamefont {P.}~\bibnamefont
  {Zoller}},\ }\bibinfo {title} {Quantum markov processes},\ in\ \href@noop {}
  {\emph {\bibinfo {booktitle} {Quantum Noise}}}\ (\bibinfo  {publisher}
  {Springer},\ \bibinfo {year} {2004})\BibitemShut {NoStop}%
\bibitem [{\citenamefont {Galvez}\ \emph {et~al.}(2011)\citenamefont {Galvez},
  \citenamefont {Coyle}, \citenamefont {Johnson},\ and\ \citenamefont
  {Reschovsky}}]{Galvez_2011}%
  \BibitemOpen
  \bibfield  {author} {\bibinfo {author} {\bibfnamefont {E.~J.}\ \bibnamefont
  {Galvez}}, \bibinfo {author} {\bibfnamefont {L.~E.}\ \bibnamefont {Coyle}},
  \bibinfo {author} {\bibfnamefont {E.}~\bibnamefont {Johnson}},\ and\ \bibinfo
  {author} {\bibfnamefont {B.~J.}\ \bibnamefont {Reschovsky}},\ }\href
  {https://doi.org/10.1088/1367-2630/13/5/053017} {\bibfield  {journal}
  {\bibinfo  {journal} {New J. Phys.}\ }\textbf {\bibinfo {volume} {13}},\
  \bibinfo {pages} {053017} (\bibinfo {year} {2011})}\BibitemShut {NoStop}%
\bibitem [{\citenamefont {Gerry}\ and\ \citenamefont
  {Knight}(2004)}]{Gerry_Knight_2004}%
  \BibitemOpen
  \bibfield  {author} {\bibinfo {author} {\bibfnamefont {C.}~\bibnamefont
  {Gerry}}\ and\ \bibinfo {author} {\bibfnamefont {P.}~\bibnamefont {Knight}},\
  }\bibinfo {title} {Beam splitters and interferometers},\ in\ \href@noop {}
  {\emph {\bibinfo {booktitle} {Introductory Quantum Optics}}}\ (\bibinfo
  {publisher} {Cambridge University Press},\ \bibinfo {year} {2004})\ p.\
  \bibinfo {pages} {135–149}\BibitemShut {NoStop}%
\bibitem [{\citenamefont {Lombardi}\ \emph {et~al.}(2002)\citenamefont
  {Lombardi}, \citenamefont {Sciarrino}, \citenamefont {Popescu},\ and\
  \citenamefont {De~Martini}}]{PhysRevLett.88.070402}%
  \BibitemOpen
  \bibfield  {author} {\bibinfo {author} {\bibfnamefont {E.}~\bibnamefont
  {Lombardi}}, \bibinfo {author} {\bibfnamefont {F.}~\bibnamefont {Sciarrino}},
  \bibinfo {author} {\bibfnamefont {S.}~\bibnamefont {Popescu}},\ and\ \bibinfo
  {author} {\bibfnamefont {F.}~\bibnamefont {De~Martini}},\ }\href
  {https://doi.org/10.1103/PhysRevLett.88.070402} {\bibfield  {journal}
  {\bibinfo  {journal} {Phys. Rev. Lett.}\ }\textbf {\bibinfo {volume} {88}},\
  \bibinfo {pages} {070402} (\bibinfo {year} {2002})}\BibitemShut {NoStop}%
\bibitem [{\citenamefont {Lee}\ and\ \citenamefont
  {Kim}(2000)}]{PhysRevA.63.012305}%
  \BibitemOpen
  \bibfield  {author} {\bibinfo {author} {\bibfnamefont {H.-W.}\ \bibnamefont
  {Lee}}\ and\ \bibinfo {author} {\bibfnamefont {J.}~\bibnamefont {Kim}},\
  }\href {https://doi.org/10.1103/PhysRevA.63.012305} {\bibfield  {journal}
  {\bibinfo  {journal} {Phys. Rev. A}\ }\textbf {\bibinfo {volume} {63}},\
  \bibinfo {pages} {012305} (\bibinfo {year} {2000})}\BibitemShut {NoStop}%
\bibitem [{\citenamefont {Knill}\ \emph {et~al.}(2001)\citenamefont {Knill},
  \citenamefont {Laflamme},\ and\ \citenamefont
  {Milburn}}]{knill_milburn_linearoptics_2001}%
  \BibitemOpen
  \bibfield  {author} {\bibinfo {author} {\bibfnamefont {E.}~\bibnamefont
  {Knill}}, \bibinfo {author} {\bibfnamefont {R.}~\bibnamefont {Laflamme}},\
  and\ \bibinfo {author} {\bibfnamefont {G.~J.}\ \bibnamefont {Milburn}},\
  }\href {https://doi.org/https://doi.org/10.1038/35051009} {\bibfield
  {journal} {\bibinfo  {journal} {Nature}\ }\textbf {\bibinfo {volume} {409}},\
  \bibinfo {pages} {46} (\bibinfo {year} {2001})}\BibitemShut {NoStop}%
\bibitem [{\citenamefont {Huang}(2015)}]{huang_quantum_2015}%
  \BibitemOpen
  \bibfield  {author} {\bibinfo {author} {\bibfnamefont {S.}~\bibnamefont
  {Huang}},\ }\href {https://doi.org/10.1103/PhysRevA.92.043845} {\bibfield
  {journal} {\bibinfo  {journal} {Phys. Rev. A}\ }\textbf {\bibinfo {volume}
  {92}},\ \bibinfo {pages} {043845} (\bibinfo {year} {2015})}\BibitemShut
  {NoStop}%
\bibitem [{\citenamefont {Massar}\ and\ \citenamefont
  {Popescu}(1995)}]{PhysRevLett.74.1259}%
  \BibitemOpen
  \bibfield  {author} {\bibinfo {author} {\bibfnamefont {S.}~\bibnamefont
  {Massar}}\ and\ \bibinfo {author} {\bibfnamefont {S.}~\bibnamefont
  {Popescu}},\ }\href {https://doi.org/10.1103/PhysRevLett.74.1259} {\bibfield
  {journal} {\bibinfo  {journal} {Phys. Rev. Lett.}\ }\textbf {\bibinfo
  {volume} {74}},\ \bibinfo {pages} {1259} (\bibinfo {year}
  {1995})}\BibitemShut {NoStop}%
\bibitem [{\citenamefont {Hammerer}\ \emph {et~al.}(2005)\citenamefont
  {Hammerer}, \citenamefont {Wolf}, \citenamefont {Polzik},\ and\ \citenamefont
  {Cirac}}]{PhysRevLett.94.150503}%
  \BibitemOpen
  \bibfield  {author} {\bibinfo {author} {\bibfnamefont {K.}~\bibnamefont
  {Hammerer}}, \bibinfo {author} {\bibfnamefont {M.~M.}\ \bibnamefont {Wolf}},
  \bibinfo {author} {\bibfnamefont {E.~S.}\ \bibnamefont {Polzik}},\ and\
  \bibinfo {author} {\bibfnamefont {J.~I.}\ \bibnamefont {Cirac}},\ }\href
  {https://doi.org/10.1103/PhysRevLett.94.150503} {\bibfield  {journal}
  {\bibinfo  {journal} {Phys. Rev. Lett.}\ }\textbf {\bibinfo {volume} {94}},\
  \bibinfo {pages} {150503} (\bibinfo {year} {2005})}\BibitemShut {NoStop}%
\bibitem [{\citenamefont {G\"undo\ifmmode~\breve{g}\else \u{g}\fi{}an}\ \emph
  {et~al.}(2012)\citenamefont {G\"undo\ifmmode~\breve{g}\else \u{g}\fi{}an},
  \citenamefont {Ledingham}, \citenamefont {Almasi}, \citenamefont
  {Cristiani},\ and\ \citenamefont {de~Riedmatten}}]{PhysRevLett.108.190504}%
  \BibitemOpen
  \bibfield  {author} {\bibinfo {author} {\bibfnamefont {M.}~\bibnamefont
  {G\"undo\ifmmode~\breve{g}\else \u{g}\fi{}an}}, \bibinfo {author}
  {\bibfnamefont {P.~M.}\ \bibnamefont {Ledingham}}, \bibinfo {author}
  {\bibfnamefont {A.}~\bibnamefont {Almasi}}, \bibinfo {author} {\bibfnamefont
  {M.}~\bibnamefont {Cristiani}},\ and\ \bibinfo {author} {\bibfnamefont
  {H.}~\bibnamefont {de~Riedmatten}},\ }\href
  {https://doi.org/10.1103/PhysRevLett.108.190504} {\bibfield  {journal}
  {\bibinfo  {journal} {Phys. Rev. Lett.}\ }\textbf {\bibinfo {volume} {108}},\
  \bibinfo {pages} {190504} (\bibinfo {year} {2012})}\BibitemShut {NoStop}%
\bibitem [{\citenamefont {Mair}\ \emph {et~al.}(2001)\citenamefont {Mair},
  \citenamefont {Vaziri}, \citenamefont {Weihs},\ and\ \citenamefont
  {Zeilinger}}]{mair_entanglement_2001}%
  \BibitemOpen
  \bibfield  {author} {\bibinfo {author} {\bibfnamefont {A.}~\bibnamefont
  {Mair}}, \bibinfo {author} {\bibfnamefont {A.}~\bibnamefont {Vaziri}},
  \bibinfo {author} {\bibfnamefont {G.}~\bibnamefont {Weihs}},\ and\ \bibinfo
  {author} {\bibfnamefont {A.}~\bibnamefont {Zeilinger}},\ }\href
  {https://doi.org/10.1038/35085529} {\bibfield  {journal} {\bibinfo  {journal}
  {Nature}\ }\textbf {\bibinfo {volume} {412}},\ \bibinfo {pages} {313}
  (\bibinfo {year} {2001})}\BibitemShut {NoStop}%
\bibitem [{\citenamefont {Kwiat}\ \emph {et~al.}(1995)\citenamefont {Kwiat},
  \citenamefont {Mattle}, \citenamefont {Weinfurter}, \citenamefont
  {Zeilinger}, \citenamefont {Sergienko},\ and\ \citenamefont
  {Shih}}]{kwiat_new_1995}%
  \BibitemOpen
  \bibfield  {author} {\bibinfo {author} {\bibfnamefont {P.~G.}\ \bibnamefont
  {Kwiat}}, \bibinfo {author} {\bibfnamefont {K.}~\bibnamefont {Mattle}},
  \bibinfo {author} {\bibfnamefont {H.}~\bibnamefont {Weinfurter}}, \bibinfo
  {author} {\bibfnamefont {A.}~\bibnamefont {Zeilinger}}, \bibinfo {author}
  {\bibfnamefont {A.~V.}\ \bibnamefont {Sergienko}},\ and\ \bibinfo {author}
  {\bibfnamefont {Y.}~\bibnamefont {Shih}},\ }\href
  {https://doi.org/10.1103/PhysRevLett.75.4337} {\bibfield  {journal} {\bibinfo
   {journal} {Phys. Rev. Lett.}\ }\textbf {\bibinfo {volume} {75}},\ \bibinfo
  {pages} {4337} (\bibinfo {year} {1995})}\BibitemShut {NoStop}%
\bibitem [{\citenamefont {Krenn}\ \emph {et~al.}(2017)\citenamefont {Krenn},
  \citenamefont {Malik}, \citenamefont {Erhard},\ and\ \citenamefont
  {Zeilinger}}]{doi:10.1098/rsta.2015.0442}%
  \BibitemOpen
  \bibfield  {author} {\bibinfo {author} {\bibfnamefont {M.}~\bibnamefont
  {Krenn}}, \bibinfo {author} {\bibfnamefont {M.}~\bibnamefont {Malik}},
  \bibinfo {author} {\bibfnamefont {M.}~\bibnamefont {Erhard}},\ and\ \bibinfo
  {author} {\bibfnamefont {A.}~\bibnamefont {Zeilinger}},\ }\href
  {https://doi.org/10.1098/rsta.2015.0442} {\bibfield  {journal} {\bibinfo
  {journal} {Philos. Trans. R. Soc. A: Mathematical, Physical and Engineering
  Sciences}\ }\textbf {\bibinfo {volume} {375}},\ \bibinfo {pages} {20150442}
  (\bibinfo {year} {2017})}\BibitemShut {NoStop}%
\bibitem [{\citenamefont {Kiesewetter}\ \emph {et~al.}(2017)\citenamefont
  {Kiesewetter}, \citenamefont {Teh}, \citenamefont {Drummond},\ and\
  \citenamefont {Reid}}]{kiesewetter_pulsed_2017}%
  \BibitemOpen
  \bibfield  {author} {\bibinfo {author} {\bibfnamefont {S.}~\bibnamefont
  {Kiesewetter}}, \bibinfo {author} {\bibfnamefont {R.}~\bibnamefont {Teh}},
  \bibinfo {author} {\bibfnamefont {P.}~\bibnamefont {Drummond}},\ and\
  \bibinfo {author} {\bibfnamefont {M.}~\bibnamefont {Reid}},\ }\href
  {https://doi.org/10.1103/PhysRevLett.119.023601} {\bibfield  {journal}
  {\bibinfo  {journal} {Phys. Rev. Lett.}\ }\textbf {\bibinfo {volume} {119}},\
  \bibinfo {pages} {023601} (\bibinfo {year} {2017})}\BibitemShut {NoStop}%
\bibitem [{\citenamefont {Ghobadi}\ \emph {et~al.}(2011)\citenamefont
  {Ghobadi}, \citenamefont {Bahrampour},\ and\ \citenamefont
  {Simon}}]{GhobadiPRA2011}%
  \BibitemOpen
  \bibfield  {author} {\bibinfo {author} {\bibfnamefont {R.}~\bibnamefont
  {Ghobadi}}, \bibinfo {author} {\bibfnamefont {A.~R.}\ \bibnamefont
  {Bahrampour}},\ and\ \bibinfo {author} {\bibfnamefont {C.}~\bibnamefont
  {Simon}},\ }\href {https://doi.org/10.1103/PhysRevA.84.033846} {\bibfield
  {journal} {\bibinfo  {journal} {Phys. Rev. A}\ }\textbf {\bibinfo {volume}
  {84}},\ \bibinfo {pages} {033846} (\bibinfo {year} {2011})}\BibitemShut
  {NoStop}%
\bibitem [{\citenamefont {Horodecki}\ \emph {et~al.}(1999)\citenamefont
  {Horodecki}, \citenamefont {Horodecki},\ and\ \citenamefont
  {Horodecki}}]{PhysRevA.60.1888}%
  \BibitemOpen
  \bibfield  {author} {\bibinfo {author} {\bibfnamefont {M.}~\bibnamefont
  {Horodecki}}, \bibinfo {author} {\bibfnamefont {P.}~\bibnamefont
  {Horodecki}},\ and\ \bibinfo {author} {\bibfnamefont {R.}~\bibnamefont
  {Horodecki}},\ }\href {https://doi.org/10.1103/PhysRevA.60.1888} {\bibfield
  {journal} {\bibinfo  {journal} {Phys. Rev. A}\ }\textbf {\bibinfo {volume}
  {60}},\ \bibinfo {pages} {1888} (\bibinfo {year} {1999})}\BibitemShut
  {NoStop}%
\bibitem [{\citenamefont {Johansson}\ \emph {et~al.}(2012)\citenamefont
  {Johansson}, \citenamefont {Nation},\ and\ \citenamefont {Nori}}]{qutip}%
  \BibitemOpen
  \bibfield  {author} {\bibinfo {author} {\bibfnamefont {J.}~\bibnamefont
  {Johansson}}, \bibinfo {author} {\bibfnamefont {P.}~\bibnamefont {Nation}},\
  and\ \bibinfo {author} {\bibfnamefont {F.}~\bibnamefont {Nori}},\ }\href
  {https://doi.org/10.1016/j.cpc.2012.02.021} {\bibfield  {journal} {\bibinfo
  {journal} {Computer Physics Communications}\ }\textbf {\bibinfo {volume}
  {183}},\ \bibinfo {pages} {1760} (\bibinfo {year} {2012})}\BibitemShut
  {NoStop}%
\bibitem [{\citenamefont {Bouillard}\ \emph {et~al.}(2019)\citenamefont
  {Bouillard}, \citenamefont {Boucher}, \citenamefont {Ferrer~Ortas},
  \citenamefont {Pointard},\ and\ \citenamefont
  {Tualle-Brouri}}]{PhysRevLett.122.210501}%
  \BibitemOpen
  \bibfield  {author} {\bibinfo {author} {\bibfnamefont {M.}~\bibnamefont
  {Bouillard}}, \bibinfo {author} {\bibfnamefont {G.}~\bibnamefont {Boucher}},
  \bibinfo {author} {\bibfnamefont {J.}~\bibnamefont {Ferrer~Ortas}}, \bibinfo
  {author} {\bibfnamefont {B.}~\bibnamefont {Pointard}},\ and\ \bibinfo
  {author} {\bibfnamefont {R.}~\bibnamefont {Tualle-Brouri}},\ }\href
  {https://doi.org/10.1103/PhysRevLett.122.210501} {\bibfield  {journal}
  {\bibinfo  {journal} {Phys. Rev. Lett.}\ }\textbf {\bibinfo {volume} {122}},\
  \bibinfo {pages} {210501} (\bibinfo {year} {2019})}\BibitemShut {NoStop}%
\end{thebibliography}%
	

\end{document}